\documentclass[%
 reprint,
 amsmath,amssymb,
 aps,
]{revtex4-2}

\usepackage{graphicx}% Include figure files
\usepackage{dcolumn}% Align table columns on decimal point
\usepackage{bm}% bold math
\begin{document}

\preprint{APS/123-QED}

\title{Characterising arbitrary dark solitons in trapped one-dimensional Bose-Einstein condensates}% Force line breaks with \\

\author{H. A. J. Middleton-Spencer}
 \email{h.a.j.middleton-spencer2@newcastle.ac.uk}%Lines break automatically or can be forced with \\
\author{N. G. Parker}%
\author{L. Galantucci}
\author{C. F. Barenghi}
\affiliation{%
  Joint Quantum Centre (JQC) Durham–-Newcastle,
  School of Mathematics, Statistics and Physics, Newcastle University,
  Newcastle upon Tyne, NE1 7RU, United Kingdom
}%


\date{\today}% It is always \today, today,

\begin{abstract}\noindent
We present a method to detect the presence and depth of dark solitons within repulsive one-dimensional harmonically trapped Bose-Einstein condensates. For a system with one soliton, we prove that the shift of the density in Fourier space directly maps onto the depth of the soliton. For multi-soliton systems, combining our spectral method with
established imaging techniques, the character of the solitons present in the condensate can be determined. We verify that the detection of solitons by the spectral shift works in the presence of waves induced by density engineering methods. Finally we discuss implications for vortex detection in three dimensional Bose-Einstein condensates. 
%\begin{description}
%\end{description}
\end{abstract}

\maketitle
\section{Introduction} \noindent
Quantifying the nature of turbulence in a three-dimensional (3D) Bose-Einstein condensate requires a way of identifying individual phase defects present in the system. Turbulence, characterised by a disordered system of entangled vortices and sound waves with a power law in the energy system, provides a challenge of visualisation. Such an entangled 3D system provides a difficult system to optically image; if the individual vortices are not aligned with the direction of visualisation then they are barely visible within the turbulent cloud. We wish to provide a quantitative method of detecting vortices within 3D condensate. We hence take a step back and begin our journey with the study of one-dimensional (1D) repulsive Bose-Einstein Condensates (BECs) and the dark solitons (the 1D analogue for vortices for a repulsive condensate) within them. Solitons, localised waves characterised by the balance of nonlinearity and dispersion, are ubiquitous in nature: they are present in many nonlinear systems, from optics \cite{kivshar1998,kivshar1989soliton,kibler2010peregrine}, thin films \cite{chen1993microwave,drozdovskii2010formation} and fluids \cite{camassa1993integrable,kodama2010kp,PhysRevLett.110.124101} to atomic BECs \cite{PhysRevLett.83.5198,Science.287.97,Frantzeskakis2010,Science.293.663,PhysRevLett.98.180401,PhysRevLett.99.160405,NatPhys.4.496,PhysRevLett.101.170404,PhysRevLett.101.120406,PhysRevLett.101.130401,PhysRevLett.125.030402,PhysRevLett.125.030401,PNAS.114.2503,PhysRevLett.119.150403,PhysRevA.101.053629}.
Multiple methods of creating dark solitons 
\cite{PhysRevLett.80.2972,PhysRevA.63.051601,Scott_1998} 
were proposed shortly after realisation of BECs \cite{Anderson198,PhysRevLett.75.3969,PhysRevLett.75.1687,PhysRevLett.79.1170} and the subsequent creation of low dimension condensates \cite{PhysRevLett.87.130402}. These methods can be broken into two categories: phase imprinting \cite{PhysRevA.64.043601,PhysRevLett.83.5198,Science.287.97,NatPhys.4.496} and density engineering \cite{PhysRevLett.101.130401,Scott_1998,PhysRevLett.101.170404}, and also a combination of the two  \cite{PhysRevA.63.051601,PhysRevA.101.053629}. Recently, new methods to create solitons \cite{OnDemandPrePrint}
through quenching have also been discussed \cite{halperin2020quench,PhysRevA.100.023613}. 
Dark solitons are stable in BECs confined in quasi-1D geometries, at $T=0$, and under certain specific forms of trapping potential.  If any of these constraints are broken, the solitons are prone to decay, due to unstable excitation of the dark soliton into vortex rings/pairs (the snake instability) \cite{Science.293.663,PhysRevA.76.045601,PhysRevA.60.R2665,JPhysB.33.3983,PhysRevA.62.053606,PhysRevLett.86.2926,PhysRevA.94.013609}, thermal dissipation \cite{PhysRevA.60.3220,PhysRevLett.89.110401,PhysRevA.75.051601}, and net sound emission \cite{PhysRevLett.84.2298,PhysRevLett.90.220401,PhysRevA.81.033606,PhysRevA.95.013628}, respectively.
\noindent
The first generation of dark soliton experiments in BECs showed the possibility of non-dispersive solitary waves propagating through the background condensate \cite{PhysRevA.64.043601,PhysRevLett.83.5198,Science.287.97,Anderson198,Science.293.663}. In these particular experiments, propagation was very short-lived. Dark solitons were shown to break down quickly due to a mixture of thermodynamic and dynamical instabilities which occurred if the background density was not strongly enough within the quasi-1D regime \cite{PhysRevA.60.R2665}. Unlike theoretical work based on the 1D mean field limit, experimental quasi-1D condensates 
consist of cigar-like clouds of trapped atoms in which a strong confinement prevents excitations in the radial plane. The second generation of experiments was able to both keep the condensate cold and confined enough in the radial directions to sustain dark solitons for long periods (that is, for at least one oscillation of the soliton) \cite{PhysRevLett.101.130401,NatPhys.4.496}. Notable experiments include verification of the oscillation frequency of the soliton in a harmonic trap $\omega/\sqrt{2}$ (where $\omega$ is the trapping frequency of the 1D system) for systems consisting of a single soliton \cite{shomroni2009evidence} and the deviation from this prediction for multiple soliton systems \cite{PhysRevLett.101.130401}. Interesting theoretical models include the study of interactions of multiple dark solitons \cite{PhysRevA.100.033607,PhysRevA.81.063604} and solitons in two-component systems \cite{PhysRevA.84.053630,middelkamp2011dynamics,PhysRevA.100.013626}. 
\noindent 
Analysis of the complex structure present in a condensate - be that a single vortex/ soliton or quantum turbulence \cite{barenghi2014introduction,tsatsos2016quantum,PhysRevLett.103.045301,Nature.539.72} - requires accurate detection of the density. Techniques to
image the density of the condensate include dispersive methods \cite{Andrews84}, absorption \cite{doi:10.1063/1.4747163,absorption} and phase-contrast imaging \cite{PhysRevLett.78.985,PhysRevA.81.053632}. Each technique has its own advantages and disadvantages; the choice is made depending on the type of experiment taking place. 
Imaging happens either \textit{in situ} (that is, imaging the trapped condensate) or during a time-of-flight (TOF) expansion (the release of the trap). Absorption methods, either \textit{in situ} or after TOF expansion, are inherently destructive -  heating by the imaging laser and loss of the condensate in the case of TOF. The viability \cite{PhysRevLett.93.180402} and search for minimally/ non-destructive imaging techniques has been an active topic recently \cite{Gauthier:s,PhysRevA.91.023621,Shin_2D}. A particularly effective method of imaging the condensate with minimal destruction of the sample called Partial Transfer Absorption Imaging (PTAI) was experimentally demonstrated by Freilich \textit{et al} \cite{Freilich1182} and perfected by Ramanathan \textit{et al} \cite{doi:10.1063/1.4747163} and has the distinct advantage of working for almost any optical depth. 
A small given percentage of atoms are outcoupled and imaged, leaving the original condensate almost unaltered. In this way the condensate can be imaged up to 50 times before breakdown \cite{Seroka:19}; The PTAI method was used successfully to visualise solitonic vortices \cite{Donadello2014} and reconnecting vortex lines \cite{PhysRevLett.115.170402,Serafini2017} in cigar shaped condensates. PTAI can be particular useful in studying the evolving dynamics of moving solitons within a condensate. Shining light upon the outcoupled cloud and projecting this onto a charge-coupled device camera, experimentalists can gain  and thus multiple column integrated density profiles of the system can be gained  \cite{Hueck:17,Putra}. 
\newline\noindent
We will employ the success of the PTAI method to show that, by taking multiple snapshots of a condensate, there is a link between the averaged density spectra and the dark solitons present. In this paper, we solve the Gross-Pitaevskii equation for a harmonically trapped 1D system with dark solitons present. We will study condensates with single and multiple solitons and introduce the idea of a spectral shift to identify the depth of any soliton present. We utilise the density engineering method in order to study more experimentally valid systems and finally, will discuss the implications for 3D condensates.
%%%%%%%%%%%%%%%%%%%%%%%%%%%

\section{Model} \noindent
In the zero temperature limit, the mean field approximation for the 
wavefunction $\Psi(x,y,z,t)$ of the condensate provides a 
quantitative model of the dynamics in the form of the Gross-Pitaevskii 
equation (GPE),
\begin{equation}
i\hbar\frac{\partial\Psi}{\partial t}= -\frac{\hbar^2}{2m}\nabla^2\Psi + V(x,y,z)\Psi + g|\Psi|^2\Psi - \mu\Psi,
\label{eq:GPE}
\end{equation}

\noindent
under the normalization $\int|\Psi|^2 d^3r=\mathcal{N}$, where $\mathcal{N}$ is the number of atoms, $\hbar$  is the reduced Planck constant $h/2\pi$, $m$ the mass of the atomic species, $\mu$ the chemical potential, $g=4\pi\hbar^2a/m$, $a$  the scattering length of the species, and $V(x,y,z)$ the trapping potential which hereafter we assume is harmonic, of the form $V=m[\omega^2_\perp(y^2+z^2)+\omega^2_xx^2]/2$ where $\omega_\perp$ and $\omega_x$ are the trap's parameters. We can reduce eq. (\ref{eq:GPE}) \cite{barenghi2016primer} by taking the perpendicular trapping frequencies to be sufficiently large, $\omega_{\perp} \gg \omega_x$, and integrate out the dependence on $y,z$. We also hence rescale the chemical potential $\mu_{1D}=\mu-\hbar\omega_\perp$ and  $g_{1D}=g/2\pi\ell^2_{\perp}$ where $\ell_{\perp}$ is the harmonic oscillator length $\ell_{\perp}=\sqrt{\hbar/m\omega_{\perp}}$. We present our results using the natural units for a homogeneous condensate, which are  time $\tau=\hbar/\mu$, length $\xi=\hbar/\sqrt{m\mu}$ and peak density $n_0=\mu/g$.  There are two limits of elongated condensates \cite{PhysRevA.66.043610}; when $an_0\gg1$ we enter the \textit{3D cigar limit}, and when $an_0\ll1$ we are in the \textit{1D mean field limit}. In this paper, we work solely in the 1D mean field limit. Written in terms of these natural units, the 1D GPE is
\begin{equation}
i\frac{\partial\Psi'}{\partial t'}=-\frac{1}{2}\frac{\partial^2 \Psi'}{\partial x'^2}+\frac{1}{2}\omega'^2x'^2\Psi'+|\Psi'|^2\Psi'-\Psi'
\end{equation}  
\noindent
where $t'=t/\tau$, $x'=x/\xi$ and $\omega'=\omega_x\tau$. All results presented are for $\omega'=0.02$. We normalise the condensate $\int \vert \Psi' \vert^2 dx'=\mathcal{N}'$ so that the peak density (the density at the trap minimum) $n_0'$ is unity. For a condensate with $\omega'=0.02$, this corresponds to $\mathcal{N}'\approx 94.2$. We choose $\omega'=0.02$ to ensure we are within the Thomas-Fermi limit  $R_{x}'\gg\omega'^{(-1/2)}$. Here  $R_x'=R_x/\xi$ and $\omega'^{(-1/2)}$ are the Thomas-Fermi radius and dimensionless harmonic oscillator length of the condensate respectively. Provided this relation is satisfied, we can model the shape of the 1D condensate by the Thomas-Fermi profile \cite{PhysRevA.66.043610,PhysRevLett.76.6},
\begin{equation}
n_{TF}(x) = 
\begin{cases}
n'_0\Big(1-\frac{{x}^{'2}}{R_x^{'2}}\Big) & \text{for } |x'| \leq R'_x\\ 0 & \text{for } |x'| > R'_x
\end{cases},
\label{eq:TF}.
\end{equation}
\noindent
 For $\omega'=0.02$, $R'_{x}\approx70$.  For clarity, hereafter primes are dropped throughout.
\newline As discussed above,  phase defects in a repulsive 1D condensate take the form of dark solitons. The (dimensional) analytic expression for a dark soliton of prescribed speed $v$ in a homogeneous background is \cite{Frantzeskakis2010},
\begin{equation}
\Psi_S(x) = \sqrt{n}\Bigg[B\tanh{\Bigg(\frac{B(x-x_0)}{\xi}\Bigg)}+i\frac{v}{c}\Bigg],
\label{eq:soliton}
\end{equation}
\noindent
where $B=\sqrt{1-v^2/c^2}$, 
and $n$, $c$ and $\xi$ are density, sound speed and healing length. To set up the initial condition for a numerical simulation of a single soliton in our condensate we multiply $\Psi_s$ (cast in dimensionless form) by the ground state $\Psi$ of a
harmonically trapped condensate. The density depletion at the minimum of the resulting condensate and the speed of the soliton are related via,
\begin{equation}
\Delta n = 1-v^2/c^2.
\end{equation}
\noindent
In the numerical simulations the GPE is solved via a fourth-order 
Runge-Kutta method using \texttt{MATLAB} with $dx=0.1$ and $dt=0.01$. 
At desired times, we compute the density spectrum, defined as the 
Fourier transform of the density of the condensate, 
$\tilde{n}(k) = \mathcal{F}(n(x))=\mathcal{F}(|\Psi(x)|^2)$,
using inbuilt \texttt{MATLAB} subroutines for the Fast Fourier Transform, where $k$ is the wavenumber $k=2\pi/x$.

\section{Ground state} 
\begin{figure}[h!]
	\centering
	{\includegraphics[trim={9.5cm 5cm 10cm 5cm},clip,width=0.35\textwidth]{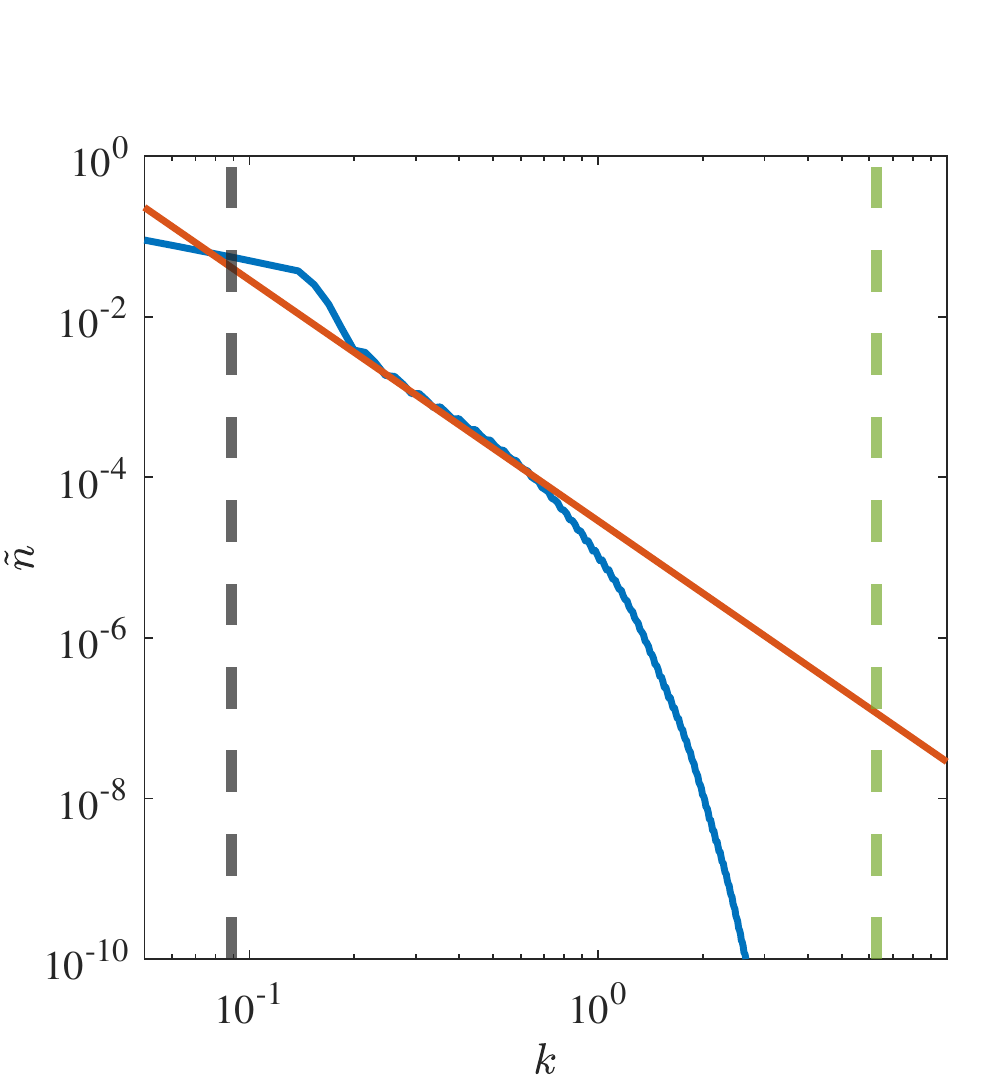}}
	\caption{The density spectra $\tilde{n}$ of the ground state ({blue}) and of its Thomas-Fermi approximation, ({red}), vs wavenumber $k$; the wavenumbers corresponding to the healing length ({green}) and the  Thomas-Fermi radius ({black}) are marked as vertical dashed lines.}
	\label{fig:ground state_solution}
\end{figure}

\noindent
We begin our investigation by first studying the density spectrum
of the ground state of the 1D harmonically trapped condensate. 
Figure \ref{fig:ground state_solution} compares the density spectrum of 
the ground state solution, obtained numerically, to the density spectrum of its Thomas-Fermi approximation.
The two spectra are consistent in the region of k-space corresponding 
to the central region within the Thomas-Fermi width, $R_x$, and the healing length, $\xi$, but they deviate at large wavenumbers. This is expected: whereas the edges of the density profile of the ground state taper off smoothly, the abrupt cut-off of the Thomas-Fermi profile means that the spectrum has relatively large power at large $k$. 

\section{Single Soliton} 
\begin{figure}[h!]
 	\centering
 	{\includegraphics[trim={4.2cm 11cm 4.2cm 11.5cm},clip,width=0.49\textwidth]{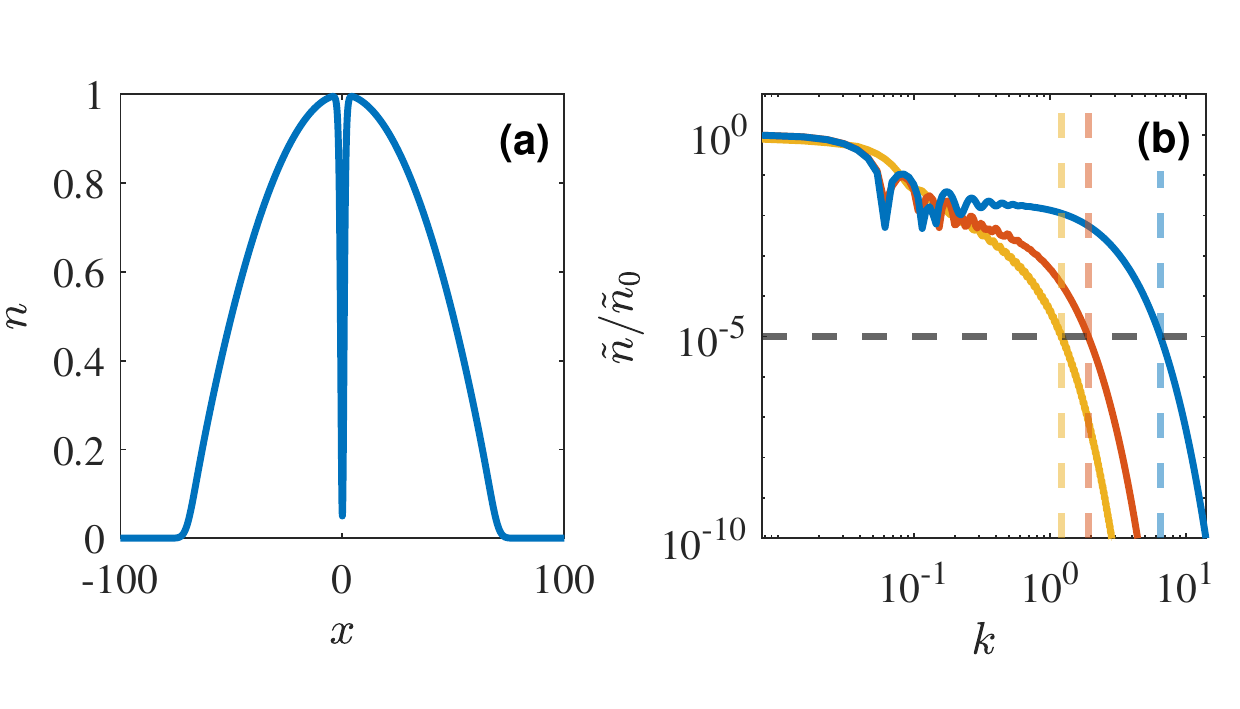}}
 	\caption{(a) The density of a condensate with a soliton with a depth $\Delta n =0.9$ inserted at the centre; ({b}) the scaled spectra $\tilde{n}/\tilde{n}_0$ of a ground state condensate ({yellow}) and of a condensate containing a soliton of depth $\Delta n=0.1$ ({red}) and a soliton of depth $\Delta n=0.9$ ({blue}), with $n_i$ marked by horizontal black line. The corresponding intercept wavenumber $k_i$s are also marked by vertical dashed lines.}
 	\label{fig:onesolitonexample}
 \end{figure}

\begin{figure}[h!]
	\centering
	{\includegraphics[trim={6cm 11.5cm 6cm 11.5cm},clip,width=0.3\textwidth]{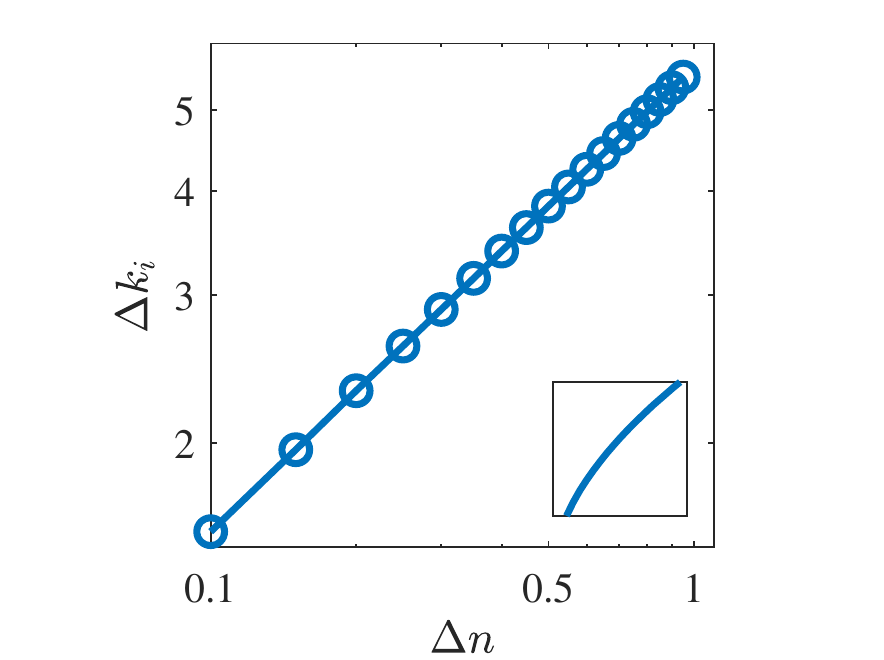}}
\caption{The spectral shift $\Delta k_i$ for different soliton 
depths $\Delta n$ on a logarithmic scale ({main fig.}) and linear scale 
({inset; same ranges as the main fig.}). }
	\label{fig:onesolitonnd}
\end{figure}

\noindent
As discussed above, a condensate containing a single soliton is easily obtained by multiplying the ground state wavefunction by the expression for a dark soliton in a homogeneous system, eq. (\ref{eq:soliton}), (see fig. \ref{fig:onesolitonexample}(a)). Upon comparing the spectrum of the ground state and the spectrum of the single-soliton condensate, we notice a shift rightwards towards the 
smaller length scale (larger k). We quantify this spectral shift and relate it to the soliton's depth $\Delta n$ (hence its speed $v$), we proceed in the following way.
\noindent
Consider a single soliton at the condensate centre. As mentioned earlier, a single soliton in a harmonically trapped condensate oscillates around the trap's minimum at frequency $\omega/\sqrt{2}$ with an amplitude depending on the soliton depth $\Delta n$. We let the system evolve in time for $t=500$, averaging density spectra taken
for every $0.5$ time units. We verify that the the resulting time-averaged spectrum does not change in time. Figure \ref{fig:onesolitonexample}(b) shows that the addition of a soliton drastically shifts the density spectrum to larger wavenumbers (compare the blue curve with the red and yellow curves). For the sake of making comparisons, the density spectra $\tilde{n}(k)$ we present are rescaled by $\tilde{n}_0$, defined as $\tilde{n}$ for $k\rightarrow0$. To quantify the shift of density spectrum to larger wavenumbers arising from the presence of solitons, we define the relative spectral shift $\Delta k_i=k_i/k_{i}^{(0)}$; 
here $k_i$, which we refer to as the intercept wavenumber,
is the wavenumber corresponding to the value
$\tilde{n}_i = 10^{-5}\tilde{n}_0$ in the presence of the soliton,
and $k_i^{(0)}$ is the intercept wavenumber of the 
ground state.  For the ground state with $\omega=0.02$, $k_i^{(0)}=1.22$. Comparing $\Delta k_i$ obtained for a variety of soliton depths from $0.1$ to $0.9$ we observe the following power law relation between soliton depth and relative spectral shift of the density,
\begin{equation}
 \Delta k_i \sim \Delta n^\alpha,
\label{eq:powerlaw}
\end{equation}
\noindent
with $\alpha=0.55$ (see fig. \ref{fig:onesolitonnd}). Clearly, the deeper the soliton the larger the spectral shift. We have verified that this result does not depend on the precise definition of $n_i$. We have also checked that eq. \ref{eq:powerlaw} is valid for a variety of harmonically trapped condensates, as long as these condensates are deeply in the Thomas-Fermi regime. Rescaling $n_0$ from 1, and hence altering the norm of the system, we have verified that the norm of the condensate does not affect the spectral shift. If we look at systems with $\omega\neq0.02$, we find that for $\omega\ll1$, the same relation (albeit with a different power law) is present. When $\omega\rightarrow1$, we see that it begins to falter; the condensate itself begins to have structure on the same lengthscales as the dark solitons.

\section{Two solitons} \noindent
The strong dependence of the spectral shift on the depth of a single soliton moving within the condensate enables us to determine the depth, and hence speed, of the soliton existing in the system. The next logical step is to assess how multiple solitons affect the density spectrum, whether a spectral shift is still observable, and finally whether it can be related to the number or the depths of the solitons.
\begin{figure}[h!]
	\centering
	{\includegraphics[trim={3cm 11.2cm 3cm 12cm},clip,width=0.49\textwidth]{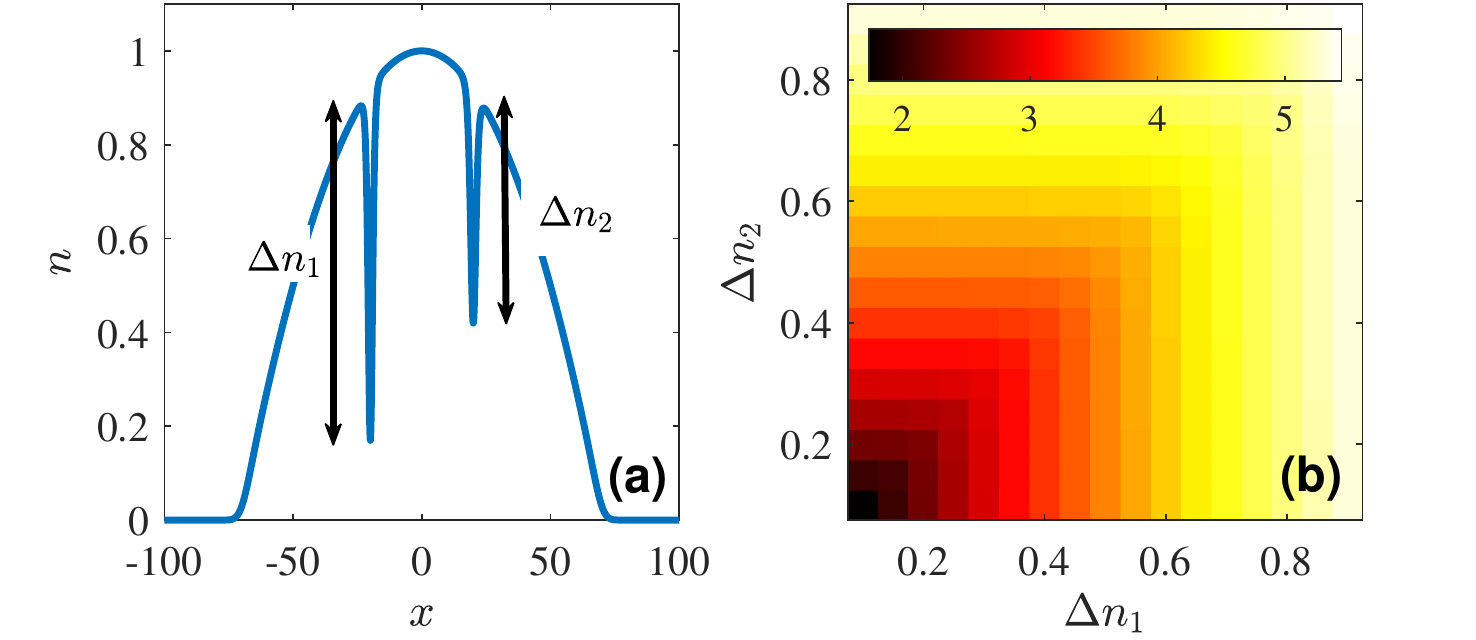}}
	\caption{(a) The density profile, $n(x)$, of condensate with two 
solitons with depths $\Delta n_{1} = 0.75$ and $\Delta n_{2} =0.5$ 
respectively; (b) the spectral shift $\Delta k_i$ (values on the
colorbar) obtained for a range of values of $\Delta n_{1}$ 
and $\Delta n_{2}$.}
	\label{fig:two_soliton_solution}
\end{figure}
\noindent
For two solitons of equal depth, the power law relationship between the soliton's depth and the relative spectral shift (eq. (\ref{eq:powerlaw})) holds true. The combination of two solitons of two different speeds, as presented in fig. \ref{fig:two_soliton_solution}(a), exhibits a more complicated spectral signature. Figure \ref{fig:two_soliton_solution}(b) shows results for two solitons. We immediately see that $\Delta k_i$ depends mainly on the deepest soliton, (the shallower soliton having only a minor effect). Although direct determination of the depths (and hence speed) is not possible from the spectra, the concentric nature of the results presented in fig. \ref{fig:two_soliton_solution}(b) shows that if we know the shift, we can easily narrow down the depths to a range of results. 
\section{Many solitons} 
\begin{figure}[h!]
	\centering
	{\includegraphics[trim={6.5cm 6.5cm 7cm 7cm},clip,width=0.49\textwidth]{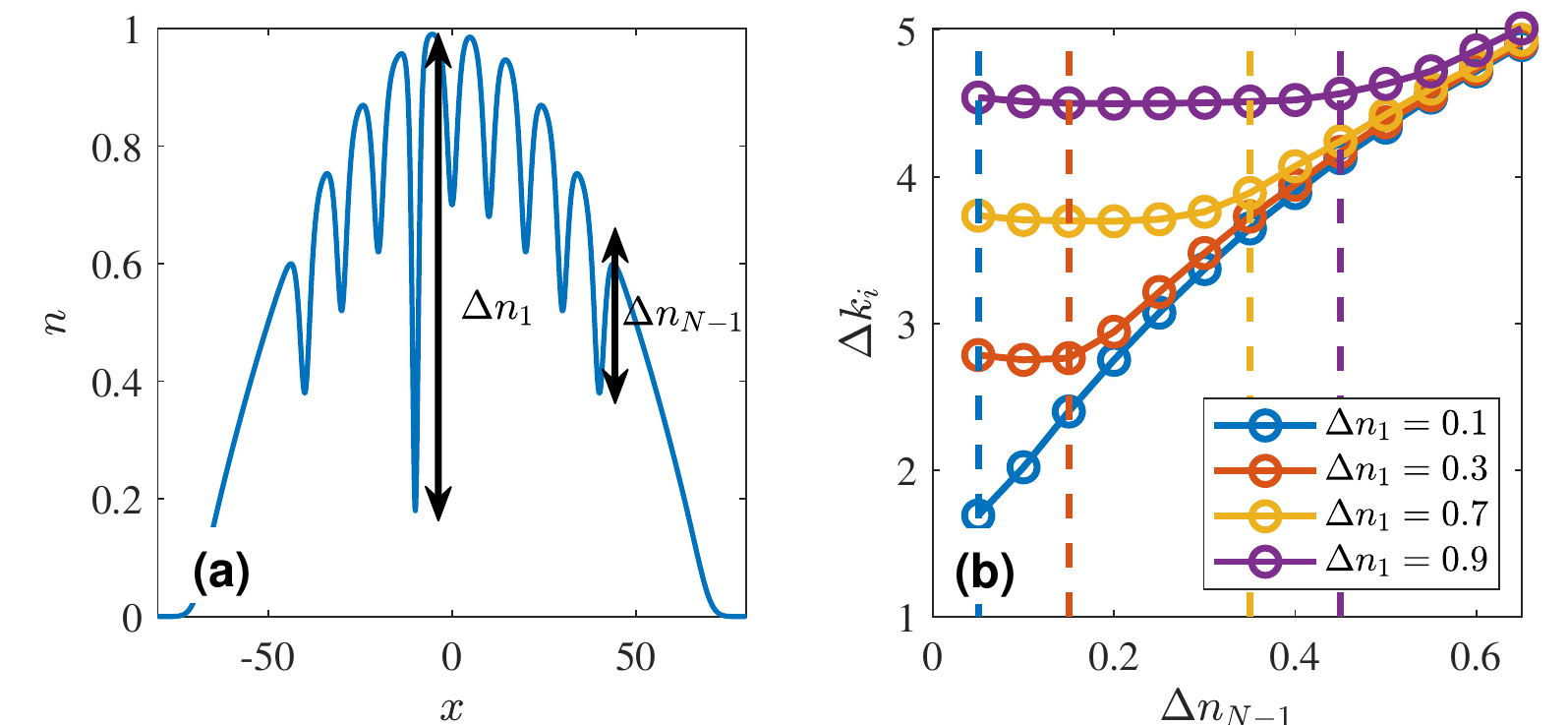}}
\caption{(a) Density of a condensate with $N=9$ solitons: 
one soliton has depth $\Delta n_{1} = 0.7$ and the other eight solitons have depth $\Delta n_{N-1}=0.3$, (b) the resulting spectral shift $\Delta k_i$ as a function of $\Delta n_{N-1}$ for varying $\Delta n_1$, with $\Delta n_1/2$ marked with vertical lines in their corresponding colours.}
	\label{fig:many_soliton_spectra}
\end{figure}

\noindent
We have established that in a two-soliton system the spectral
shift is mostly affected by the deepest soliton. However, the methods of creating solitons, as discussed in the Introduction, can often lead to a train of solitons. To make better contact with experiments, in this section we describe the spectral shift caused by a relatively large number, $N$, of solitons. We choose $N=9$, with the first soliton of depth $\Delta n_1$ and the other eight of the same depth $\Delta n_{N-1}$, as shown by fig. \ref{fig:many_soliton_spectra}(a).
\noindent
Figure \ref{fig:many_soliton_spectra} further supports the previous finding that the deepest soliton contributes the most to the shift of spectrum, hence a higher value of $\Delta k_i$. The spectral shifts shown in the figure display the same dependence on solitons' depth displayed in fig. \ref{fig:two_soliton_solution}: that is $\Delta k_i$ only begins to change when $\Delta n_{N-1}$ is comparable to $\Delta n_{1}$. More precisely, $\Delta k_i$ is noticeably affected only when $\Delta n_{N-1}$ is roughly half the value of $\Delta n_{1}$ (marked with vertical lines in fig. \ref{fig:many_soliton_spectra}(b)).
\section{Effects of perturbations} 
\begin{figure}[h!]
	\centering
	{\includegraphics[trim={2.5cm 11.0cm 3.0cm 11.0cm},clip,width=0.49\textwidth]{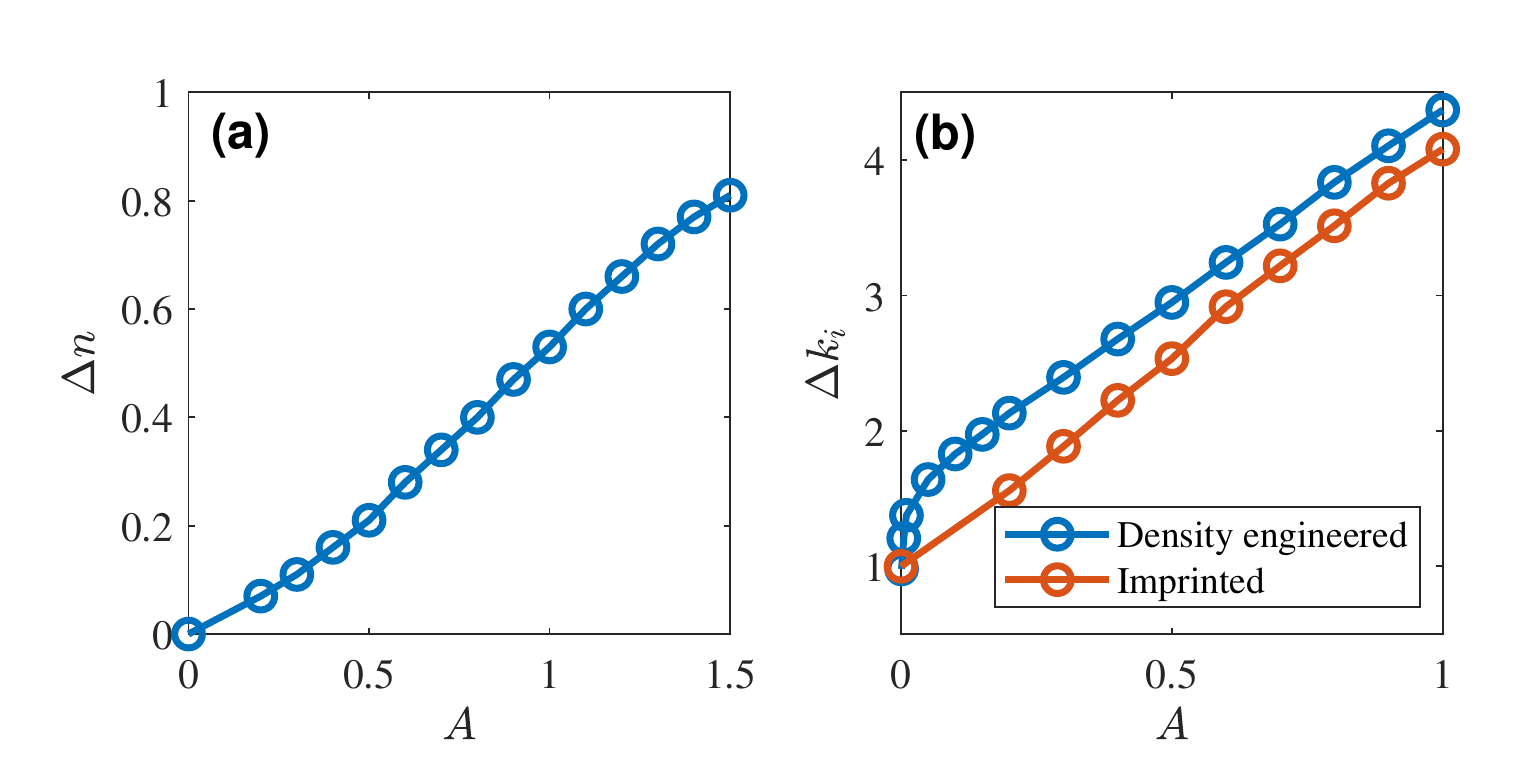}}
	\caption{(a) The depth $\Delta n$ of the leading solitons created by a Gaussian pulse with amplitude $A$;  (b) comparison between the spectral shifts $\Delta k_i$ arising from two solitons in the presence of sound waves 
(see fig. 6(a)) (blue) and two solitons of the same depth imprinted in the ground state (red).}
	\label{fig:results_gaussian_exact}
\end{figure}

\noindent
All results described in the previous sections refer to
solitons imprinted into the ground state. In many experiments, because of the method used to generate them, solitons coexist with sound waves. The following density engineering method allow us to mimic numerically this more realistic situation.
 \begin{figure}[h!]
	\centering
	{\includegraphics[trim={3.2cm 11.2cm 3.5cm 11.8cm},clip,width=0.49\textwidth]{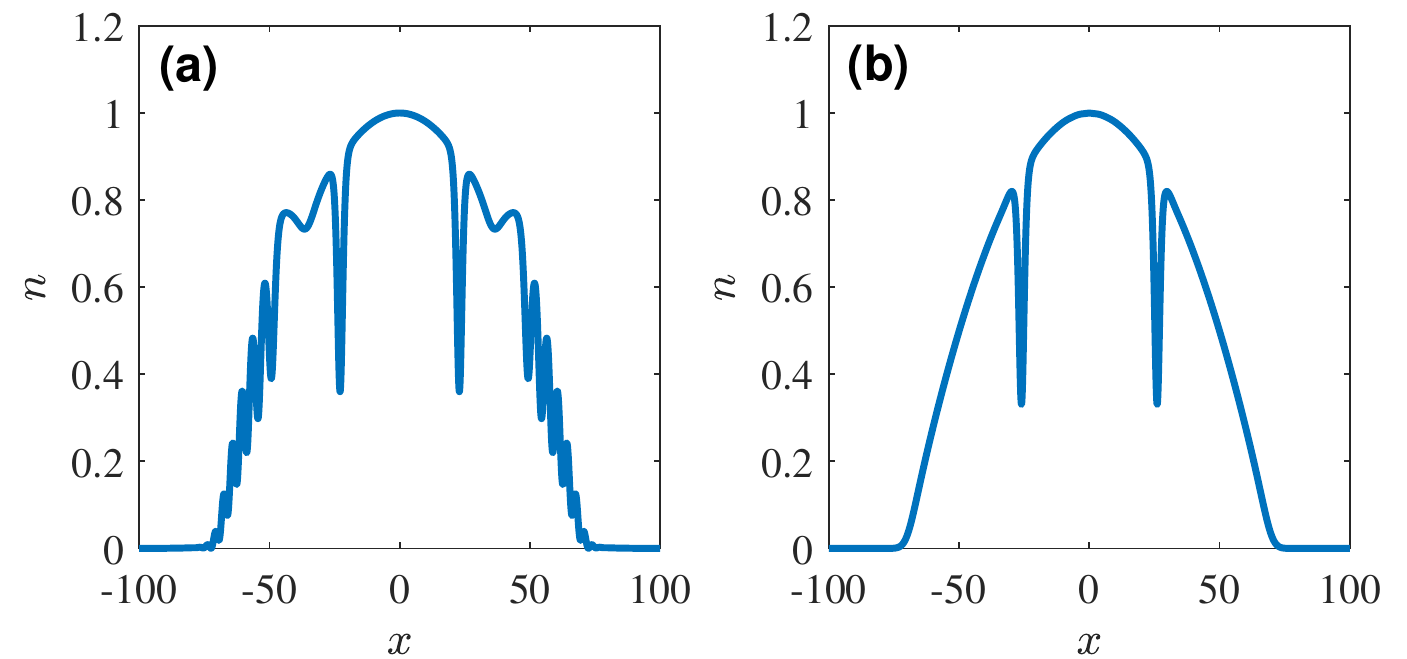}}
	\caption{(a) The density of a condensate after being excited by a pulsed Gaussian potential with $A=1$; note the creation of two main deep solitons 
(here with $\Delta n \approxeq0.53$) together with sound waves. For comparison, (b) shows two solitons of the same depth $\Delta n = 0.53$ imprinted into a ground state.}
	\label{fig:comp_gaussian_exact}
\end{figure}
\noindent
We apply a Gaussian potential of width $\sigma$ and amplitude $A$ for a time $\mathcal{T}$, before instantaneously removing it.  This pulsed Gaussian potential has the form,
\begin{equation}
V(x,t) = 
\begin{cases}
\frac{1}{2}\omega^2x^2 + A\exp{({-x^2}/{\sigma^2})}& \text{for } t \leq \mathcal{T}\\ \frac{1}{2}\omega^2x^2 & \text{for } t > \mathcal{T}
\end{cases}
\label{eq:potential}
\end{equation}
This mimics the creation of solitons via density engineering methods and causes a density dip to form at the centre of the condensate. When removed, both solitons and sound waves are generated during the collapse inwards of the two sides, as shown in fig. \ref{fig:comp_gaussian_exact}(a). We measure the depth of the two main solitons created in order to imprint them into a ground state for comparison (fig. \ref{fig:comp_gaussian_exact}(b)). We choose $\sigma=2$, $\mathcal{T}=5$ and vary the amplitude, $A$, of the central Gaussian pulse. The larger the amplitude $A$, the deeper the leading solitons created (see fig. \ref{fig:results_gaussian_exact}(a)). 
In fig. \ref{fig:results_gaussian_exact}(b), we compare the relative spectral shifts $\Delta k_i$ resulting from solitons which are imprinted in the ground state (red line) and density engineered (blue line). We see immediately that although the spectral shifts $\Delta k_i$ in the two systems are not equal, the relationship between the soliton depths and the intercept wavenumbers still holds, regardless of the presence of sound waves.

\noindent When $A\rightarrow0$, no solitons are created. Notice the rapid changes of the spectral shift for very small amplitudes $A$; these small spectral shifts correspond to systems with no detectable  soliton. The relation between the amplitude and the intercept
wavenumber $k_i$ begins to level out roughly at $A=0.2$, corresponding to two solitons of depth $\Delta n = 0.07$. 
We conclude that sound waves can shift the density spectrum, but the shift is small compared by the larger shift induced by solitons.
\section{Conclusion}
\noindent
We have presented a method for accurately ascertaining the depth (and hence the speed) of a single soliton in a harmonically trapped condensate from the density spectrum alone. We have also shown that, in a system with multiple dark solitons, the spectral shift is mainly determined by the deepest soliton. 
\newline\noindent
The analysis of the spectral shift which we have presented here for 1D systems may potentially be applied to 3D turbulent systems. A spectral shift of the momentum of the condensate has indeed already been reported in an experiment with a turbulent 3D condensate \cite{thompson2013evidence}. While in 1D phase defects are solitons whose width depends on its speed, in 3D the phase defects of the system are vortices and, since in most cases multiply charged vortices are unstable, the width of the vortex cores is constant (although, in a harmonically trapped condensate, this width increases near the edge). Therefore it may be possible to relate the measured spectral shift to the number or the length of vortices present in the system, thus providing a quantitative measure of the intensity of the turbulence in the condensate.

\begin{acknowledgments}
\noindent N. G. P., L. G. and C. F. B. acknowledge support by the Engineering and Physical Sciences Research Council (Grant No. EP/R005192/1).
\end{acknowledgments}

\bibliography{apssamp}

\providecommand{\noopsort}[1]{}\providecommand{\singleletter}[1]{#1}%
\begin{thebibliography}{78}%
\makeatletter
\providecommand \@ifxundefined [1]{%
 \@ifx{#1\undefined}
}%
\providecommand \@ifnum [1]{%
 \ifnum #1\expandafter \@firstoftwo
 \else \expandafter \@secondoftwo
 \fi
}%
\providecommand \@ifx [1]{%
 \ifx #1\expandafter \@firstoftwo
 \else \expandafter \@secondoftwo
 \fi
}%
\providecommand \natexlab [1]{#1}%
\providecommand \enquote  [1]{``#1''}%
\providecommand \bibnamefont  [1]{#1}%
\providecommand \bibfnamefont [1]{#1}%
\providecommand \citenamefont [1]{#1}%
\providecommand \href@noop [0]{\@secondoftwo}%
\providecommand \href [0]{\begingroup \@sanitize@url \@href}%
\providecommand \@href[1]{\@@startlink{#1}\@@href}%
\providecommand \@@href[1]{\endgroup#1\@@endlink}%
\providecommand \@sanitize@url [0]{\catcode `\\12\catcode `\$12\catcode
  `\&12\catcode `\#12\catcode `\^12\catcode `\_12\catcode `\%12\relax}%
\providecommand \@@startlink[1]{}%
\providecommand \@@endlink[0]{}%
\providecommand \url  [0]{\begingroup\@sanitize@url \@url }%
\providecommand \@url [1]{\endgroup\@href {#1}{\urlprefix }}%
\providecommand \urlprefix  [0]{URL }%
\providecommand \Eprint [0]{\href }%
\providecommand \doibase [0]{https://doi.org/}%
\providecommand \selectlanguage [0]{\@gobble}%
\providecommand \bibinfo  [0]{\@secondoftwo}%
\providecommand \bibfield  [0]{\@secondoftwo}%
\providecommand \translation [1]{[#1]}%
\providecommand \BibitemOpen [0]{}%
\providecommand \bibitemStop [0]{}%
\providecommand \bibitemNoStop [0]{.\EOS\space}%
\providecommand \EOS [0]{\spacefactor3000\relax}%
\providecommand \BibitemShut  [1]{\csname bibitem#1\endcsname}%
\let\auto@bib@innerbib\@empty
%</preamble>
\bibitem [{\citenamefont {Kivshar}\ and\ \citenamefont
  {Luther-Davies}(1998)}]{kivshar1998}%
  \BibitemOpen
  \bibfield  {author} {\bibinfo {author} {\bibfnamefont {Y.~S.}\ \bibnamefont
  {Kivshar}}\ and\ \bibinfo {author} {\bibfnamefont {B.}~\bibnamefont
  {Luther-Davies}},\ }\bibfield  {title} {\bibinfo {title} {Dark optical
  solitons: physics and applications},\ }\href@noop {} {\bibfield  {journal}
  {\bibinfo  {journal} {Phys. Rep.}\ }\textbf {\bibinfo {volume} {298}},\
  \bibinfo {pages} {81} (\bibinfo {year} {1998})}\BibitemShut {NoStop}%
\bibitem [{\citenamefont {Kivshar}(1989)}]{kivshar1989soliton}%
  \BibitemOpen
  \bibfield  {author} {\bibinfo {author} {\bibfnamefont {Y.~S.}\ \bibnamefont
  {Kivshar}},\ }\bibfield  {title} {\bibinfo {title} {On the soliton generation
  in optical fibres},\ }\href@noop {} {\bibfield  {journal} {\bibinfo
  {journal} {J. Phys. A}\ }\textbf {\bibinfo {volume} {22}},\ \bibinfo {pages}
  {337} (\bibinfo {year} {1989})}\BibitemShut {NoStop}%
\bibitem [{\citenamefont {Kibler}\ \emph {et~al.}(2010)\citenamefont {Kibler},
  \citenamefont {Fatome}, \citenamefont {Finot}, \citenamefont {Millot},
  \citenamefont {Dias}, \citenamefont {Genty}, \citenamefont {Akhmediev},\ and\
  \citenamefont {Dudley}}]{kibler2010peregrine}%
  \BibitemOpen
  \bibfield  {author} {\bibinfo {author} {\bibfnamefont {B.}~\bibnamefont
  {Kibler}}, \bibinfo {author} {\bibfnamefont {J.}~\bibnamefont {Fatome}},
  \bibinfo {author} {\bibfnamefont {C.}~\bibnamefont {Finot}}, \bibinfo
  {author} {\bibfnamefont {G.}~\bibnamefont {Millot}}, \bibinfo {author}
  {\bibfnamefont {F.}~\bibnamefont {Dias}}, \bibinfo {author} {\bibfnamefont
  {G.}~\bibnamefont {Genty}}, \bibinfo {author} {\bibfnamefont
  {N.}~\bibnamefont {Akhmediev}},\ and\ \bibinfo {author} {\bibfnamefont
  {J.~M.}\ \bibnamefont {Dudley}},\ }\bibfield  {title} {\bibinfo {title} {The
  peregrine soliton in nonlinear fibre optics},\ }\href@noop {} {\bibfield
  {journal} {\bibinfo  {journal} {Nat. Phys.}\ }\textbf {\bibinfo {volume}
  {6}},\ \bibinfo {pages} {790} (\bibinfo {year} {2010})}\BibitemShut {NoStop}%
\bibitem [{\citenamefont {Chen}\ \emph {et~al.}(1993)\citenamefont {Chen},
  \citenamefont {Tsankov}, \citenamefont {Nash},\ and\ \citenamefont
  {Patton}}]{chen1993microwave}%
  \BibitemOpen
  \bibfield  {author} {\bibinfo {author} {\bibfnamefont {M.}~\bibnamefont
  {Chen}}, \bibinfo {author} {\bibfnamefont {M.~A.}\ \bibnamefont {Tsankov}},
  \bibinfo {author} {\bibfnamefont {J.~M.}\ \bibnamefont {Nash}},\ and\
  \bibinfo {author} {\bibfnamefont {C.~E.}\ \bibnamefont {Patton}},\ }\bibfield
   {title} {\bibinfo {title} {Microwave magnetic-envelope dark solitons in
  yttrium iron garnet thin films},\ }\href@noop {} {\bibfield  {journal}
  {\bibinfo  {journal} {Phys. Rev. Lett.}\ }\textbf {\bibinfo {volume} {70}},\
  \bibinfo {pages} {1707} (\bibinfo {year} {1993})}\BibitemShut {NoStop}%
\bibitem [{\citenamefont {Drozdovskii}\ \emph {et~al.}(2010)\citenamefont
  {Drozdovskii}, \citenamefont {Cherkasskii}, \citenamefont {Ustinov},
  \citenamefont {Kovshikov},\ and\ \citenamefont
  {Kalinikos}}]{drozdovskii2010formation}%
  \BibitemOpen
  \bibfield  {author} {\bibinfo {author} {\bibfnamefont {A.}~\bibnamefont
  {Drozdovskii}}, \bibinfo {author} {\bibfnamefont {M.}~\bibnamefont
  {Cherkasskii}}, \bibinfo {author} {\bibfnamefont {A.~B.}\ \bibnamefont
  {Ustinov}}, \bibinfo {author} {\bibfnamefont {N.}~\bibnamefont {Kovshikov}},\
  and\ \bibinfo {author} {\bibfnamefont {B.~A.}\ \bibnamefont {Kalinikos}},\
  }\bibfield  {title} {\bibinfo {title} {Formation of envelope solitons of
  spin-wave packets propagating in thin-film magnon crystals},\ }\href@noop {}
  {\bibfield  {journal} {\bibinfo  {journal} {JETP Lett.}\ }\textbf {\bibinfo
  {volume} {91}},\ \bibinfo {pages} {16} (\bibinfo {year} {2010})}\BibitemShut
  {NoStop}%
\bibitem [{\citenamefont {Camassa}\ and\ \citenamefont
  {Holm}(1993)}]{camassa1993integrable}%
  \BibitemOpen
  \bibfield  {author} {\bibinfo {author} {\bibfnamefont {R.}~\bibnamefont
  {Camassa}}\ and\ \bibinfo {author} {\bibfnamefont {D.~D.}\ \bibnamefont
  {Holm}},\ }\bibfield  {title} {\bibinfo {title} {An integrable shallow water
  equation with peaked solitons},\ }\href@noop {} {\bibfield  {journal}
  {\bibinfo  {journal} {Phys. Rev. Lett.}\ }\textbf {\bibinfo {volume} {71}},\
  \bibinfo {pages} {1661} (\bibinfo {year} {1993})}\BibitemShut {NoStop}%
\bibitem [{\citenamefont {Kodama}(2010)}]{kodama2010kp}%
  \BibitemOpen
  \bibfield  {author} {\bibinfo {author} {\bibfnamefont {Y.}~\bibnamefont
  {Kodama}},\ }\bibfield  {title} {\bibinfo {title} {Kp solitons in shallow
  water},\ }\href@noop {} {\bibfield  {journal} {\bibinfo  {journal} {J. Phys.
  A}\ }\textbf {\bibinfo {volume} {43}},\ \bibinfo {pages} {434004} (\bibinfo
  {year} {2010})}\BibitemShut {NoStop}%
\bibitem [{\citenamefont {Chabchoub}\ \emph {et~al.}(2013)\citenamefont
  {Chabchoub}, \citenamefont {Kimmoun}, \citenamefont {Branger}, \citenamefont
  {Hoffmann}, \citenamefont {Proment}, \citenamefont {Onorato},\ and\
  \citenamefont {Akhmediev}}]{PhysRevLett.110.124101}%
  \BibitemOpen
  \bibfield  {author} {\bibinfo {author} {\bibfnamefont {A.}~\bibnamefont
  {Chabchoub}}, \bibinfo {author} {\bibfnamefont {O.}~\bibnamefont {Kimmoun}},
  \bibinfo {author} {\bibfnamefont {H.}~\bibnamefont {Branger}}, \bibinfo
  {author} {\bibfnamefont {N.}~\bibnamefont {Hoffmann}}, \bibinfo {author}
  {\bibfnamefont {D.}~\bibnamefont {Proment}}, \bibinfo {author} {\bibfnamefont
  {M.}~\bibnamefont {Onorato}},\ and\ \bibinfo {author} {\bibfnamefont
  {N.}~\bibnamefont {Akhmediev}},\ }\bibfield  {title} {\bibinfo {title}
  {Experimental observation of dark solitons on the surface of water},\
  }\href@noop {} {\bibfield  {journal} {\bibinfo  {journal} {Phys. Rev. Lett.}\
  }\textbf {\bibinfo {volume} {110}},\ \bibinfo {pages} {124101} (\bibinfo
  {year} {2013})}\BibitemShut {NoStop}%
\bibitem [{\citenamefont {Burger}\ \emph {et~al.}(1999)\citenamefont {Burger},
  \citenamefont {Bongs}, \citenamefont {Dettmer}, \citenamefont {Ertmer},
  \citenamefont {Sengstock}, \citenamefont {Sanpera}, \citenamefont
  {Shlyapnikov},\ and\ \citenamefont {Lewenstein}}]{PhysRevLett.83.5198}%
  \BibitemOpen
  \bibfield  {author} {\bibinfo {author} {\bibfnamefont {S.}~\bibnamefont
  {Burger}}, \bibinfo {author} {\bibfnamefont {K.}~\bibnamefont {Bongs}},
  \bibinfo {author} {\bibfnamefont {S.}~\bibnamefont {Dettmer}}, \bibinfo
  {author} {\bibfnamefont {W.}~\bibnamefont {Ertmer}}, \bibinfo {author}
  {\bibfnamefont {K.}~\bibnamefont {Sengstock}}, \bibinfo {author}
  {\bibfnamefont {A.}~\bibnamefont {Sanpera}}, \bibinfo {author} {\bibfnamefont
  {G.~V.}\ \bibnamefont {Shlyapnikov}},\ and\ \bibinfo {author} {\bibfnamefont
  {M.}~\bibnamefont {Lewenstein}},\ }\bibfield  {title} {\bibinfo {title} {Dark
  solitons in bose-einstein condensates},\ }\href
  {https://doi.org/10.1103/PhysRevLett.83.5198} {\bibfield  {journal} {\bibinfo
   {journal} {Phys. Rev. Lett.}\ }\textbf {\bibinfo {volume} {83}},\ \bibinfo
  {pages} {5198} (\bibinfo {year} {1999})}\BibitemShut {NoStop}%
\bibitem [{\citenamefont {Denschlag}\ \emph {et~al.}(2000)\citenamefont
  {Denschlag}, \citenamefont {Simsarian}, \citenamefont {Feder}, \citenamefont
  {Clark}, \citenamefont {Collins}, \citenamefont {Cubizolles}, \citenamefont
  {Deng}, \citenamefont {Hagley}, \citenamefont {Helmerson}, \citenamefont
  {Reinhardt} \emph {et~al.}}]{Science.287.97}%
  \BibitemOpen
  \bibfield  {author} {\bibinfo {author} {\bibfnamefont {J.}~\bibnamefont
  {Denschlag}}, \bibinfo {author} {\bibfnamefont {J.~E.}\ \bibnamefont
  {Simsarian}}, \bibinfo {author} {\bibfnamefont {D.~L.}\ \bibnamefont
  {Feder}}, \bibinfo {author} {\bibfnamefont {C.~W.}\ \bibnamefont {Clark}},
  \bibinfo {author} {\bibfnamefont {L.~A.}\ \bibnamefont {Collins}}, \bibinfo
  {author} {\bibfnamefont {J.}~\bibnamefont {Cubizolles}}, \bibinfo {author}
  {\bibfnamefont {L.}~\bibnamefont {Deng}}, \bibinfo {author} {\bibfnamefont
  {E.~W.}\ \bibnamefont {Hagley}}, \bibinfo {author} {\bibfnamefont
  {K.}~\bibnamefont {Helmerson}}, \bibinfo {author} {\bibfnamefont {W.~P.}\
  \bibnamefont {Reinhardt}}, \emph {et~al.},\ }\bibfield  {title} {\bibinfo
  {title} {Generating solitons by phase engineering of a bose-einstein
  condensate},\ }\href@noop {} {\bibfield  {journal} {\bibinfo  {journal}
  {Science}\ }\textbf {\bibinfo {volume} {287}},\ \bibinfo {pages} {97}
  (\bibinfo {year} {2000})}\BibitemShut {NoStop}%
\bibitem [{\citenamefont {Frantzeskakis}(2010)}]{Frantzeskakis2010}%
  \BibitemOpen
  \bibfield  {author} {\bibinfo {author} {\bibfnamefont {D.}~\bibnamefont
  {Frantzeskakis}},\ }\bibfield  {title} {\bibinfo {title} {Dark solitons in
  atomic bose--einstein condensates: from theory to experiments},\ }\href@noop
  {} {\bibfield  {journal} {\bibinfo  {journal} {J. Phys. A}\ }\textbf
  {\bibinfo {volume} {43}},\ \bibinfo {pages} {213001} (\bibinfo {year}
  {2010})}\BibitemShut {NoStop}%
\bibitem [{\citenamefont {Dutton}\ \emph {et~al.}(2001)\citenamefont {Dutton},
  \citenamefont {Budde}, \citenamefont {Slowe},\ and\ \citenamefont
  {Hau}}]{Science.293.663}%
  \BibitemOpen
  \bibfield  {author} {\bibinfo {author} {\bibfnamefont {Z.}~\bibnamefont
  {Dutton}}, \bibinfo {author} {\bibfnamefont {M.}~\bibnamefont {Budde}},
  \bibinfo {author} {\bibfnamefont {C.}~\bibnamefont {Slowe}},\ and\ \bibinfo
  {author} {\bibfnamefont {L.~V.}\ \bibnamefont {Hau}},\ }\bibfield  {title}
  {\bibinfo {title} {Observation of quantum shock waves created with
  ultra-compressed slow light pulses in a bose-einstein condensate},\
  }\href@noop {} {\bibfield  {journal} {\bibinfo  {journal} {Science}\ }\textbf
  {\bibinfo {volume} {293}},\ \bibinfo {pages} {663} (\bibinfo {year}
  {2001})}\BibitemShut {NoStop}%
\bibitem [{\citenamefont {Jo}\ \emph {et~al.}(2007)\citenamefont {Jo},
  \citenamefont {Choi}, \citenamefont {Christensen}, \citenamefont {Pasquini},
  \citenamefont {Lee}, \citenamefont {Ketterle},\ and\ \citenamefont
  {Pritchard}}]{PhysRevLett.98.180401}%
  \BibitemOpen
  \bibfield  {author} {\bibinfo {author} {\bibfnamefont {G.-B.}\ \bibnamefont
  {Jo}}, \bibinfo {author} {\bibfnamefont {J.-H.}\ \bibnamefont {Choi}},
  \bibinfo {author} {\bibfnamefont {C.~A.}\ \bibnamefont {Christensen}},
  \bibinfo {author} {\bibfnamefont {T.}~\bibnamefont {Pasquini}}, \bibinfo
  {author} {\bibfnamefont {Y.-R.}\ \bibnamefont {Lee}}, \bibinfo {author}
  {\bibfnamefont {W.}~\bibnamefont {Ketterle}},\ and\ \bibinfo {author}
  {\bibfnamefont {D.~E.}\ \bibnamefont {Pritchard}},\ }\bibfield  {title}
  {\bibinfo {title} {Phase-sensitive recombination of two bose-einstein
  condensates on an atom chip},\ }\href@noop {} {\bibfield  {journal} {\bibinfo
   {journal} {Phys. Rev. Lett.}\ }\textbf {\bibinfo {volume} {98}},\ \bibinfo
  {pages} {180401} (\bibinfo {year} {2007})}\BibitemShut {NoStop}%
\bibitem [{\citenamefont {Engels}\ and\ \citenamefont
  {Atherton}(2007)}]{PhysRevLett.99.160405}%
  \BibitemOpen
  \bibfield  {author} {\bibinfo {author} {\bibfnamefont {P.}~\bibnamefont
  {Engels}}\ and\ \bibinfo {author} {\bibfnamefont {C.}~\bibnamefont
  {Atherton}},\ }\bibfield  {title} {\bibinfo {title} {Stationary and
  nonstationary fluid flow of a bose-einstein condensate through a penetrable
  barrier},\ }\href {https://doi.org/10.1103/PhysRevLett.99.160405} {\bibfield
  {journal} {\bibinfo  {journal} {Phys. Rev. Lett.}\ }\textbf {\bibinfo
  {volume} {99}},\ \bibinfo {pages} {160405} (\bibinfo {year}
  {2007})}\BibitemShut {NoStop}%
\bibitem [{\citenamefont {Becker}\ \emph {et~al.}(2008)\citenamefont {Becker},
  \citenamefont {Stellmer}, \citenamefont {Soltan-Panahi}, \citenamefont
  {D{\"o}rscher}, \citenamefont {Baumert}, \citenamefont {Richter},
  \citenamefont {Kronj{\"a}ger}, \citenamefont {Bongs},\ and\ \citenamefont
  {Sengstock}}]{NatPhys.4.496}%
  \BibitemOpen
  \bibfield  {author} {\bibinfo {author} {\bibfnamefont {C.}~\bibnamefont
  {Becker}}, \bibinfo {author} {\bibfnamefont {S.}~\bibnamefont {Stellmer}},
  \bibinfo {author} {\bibfnamefont {P.}~\bibnamefont {Soltan-Panahi}}, \bibinfo
  {author} {\bibfnamefont {S.}~\bibnamefont {D{\"o}rscher}}, \bibinfo {author}
  {\bibfnamefont {M.}~\bibnamefont {Baumert}}, \bibinfo {author} {\bibfnamefont
  {E.-M.}\ \bibnamefont {Richter}}, \bibinfo {author} {\bibfnamefont
  {J.}~\bibnamefont {Kronj{\"a}ger}}, \bibinfo {author} {\bibfnamefont
  {K.}~\bibnamefont {Bongs}},\ and\ \bibinfo {author} {\bibfnamefont
  {K.}~\bibnamefont {Sengstock}},\ }\bibfield  {title} {\bibinfo {title}
  {Oscillations and interactions of dark and dark--bright solitons in
  bose--einstein condensates},\ }\href@noop {} {\bibfield  {journal} {\bibinfo
  {journal} {Nat. Phys.}\ }\textbf {\bibinfo {volume} {4}},\ \bibinfo {pages}
  {496} (\bibinfo {year} {2008})}\BibitemShut {NoStop}%
\bibitem [{\citenamefont {Chang}\ \emph {et~al.}(2008)\citenamefont {Chang},
  \citenamefont {Engels},\ and\ \citenamefont
  {Hoefer}}]{PhysRevLett.101.170404}%
  \BibitemOpen
  \bibfield  {author} {\bibinfo {author} {\bibfnamefont {J.~J.}\ \bibnamefont
  {Chang}}, \bibinfo {author} {\bibfnamefont {P.}~\bibnamefont {Engels}},\ and\
  \bibinfo {author} {\bibfnamefont {M.~A.}\ \bibnamefont {Hoefer}},\ }\bibfield
   {title} {\bibinfo {title} {Formation of dispersive shock waves by merging
  and splitting bose-einstein condensates},\ }\href
  {https://doi.org/10.1103/PhysRevLett.101.170404} {\bibfield  {journal}
  {\bibinfo  {journal} {Phys. Rev. Lett.}\ }\textbf {\bibinfo {volume} {101}},\
  \bibinfo {pages} {170404} (\bibinfo {year} {2008})}\BibitemShut {NoStop}%
\bibitem [{\citenamefont {Stellmer}\ \emph {et~al.}(2008)\citenamefont
  {Stellmer}, \citenamefont {Becker}, \citenamefont {Soltan-Panahi},
  \citenamefont {Richter}, \citenamefont {D\"orscher}, \citenamefont {Baumert},
  \citenamefont {Kronj\"ager}, \citenamefont {Bongs},\ and\ \citenamefont
  {Sengstock}}]{PhysRevLett.101.120406}%
  \BibitemOpen
  \bibfield  {author} {\bibinfo {author} {\bibfnamefont {S.}~\bibnamefont
  {Stellmer}}, \bibinfo {author} {\bibfnamefont {C.}~\bibnamefont {Becker}},
  \bibinfo {author} {\bibfnamefont {P.}~\bibnamefont {Soltan-Panahi}}, \bibinfo
  {author} {\bibfnamefont {E.-M.}\ \bibnamefont {Richter}}, \bibinfo {author}
  {\bibfnamefont {S.}~\bibnamefont {D\"orscher}}, \bibinfo {author}
  {\bibfnamefont {M.}~\bibnamefont {Baumert}}, \bibinfo {author} {\bibfnamefont
  {J.}~\bibnamefont {Kronj\"ager}}, \bibinfo {author} {\bibfnamefont
  {K.}~\bibnamefont {Bongs}},\ and\ \bibinfo {author} {\bibfnamefont
  {K.}~\bibnamefont {Sengstock}},\ }\bibfield  {title} {\bibinfo {title}
  {Collisions of dark solitons in elongated bose-einstein condensates},\ }\href
  {https://doi.org/10.1103/PhysRevLett.101.120406} {\bibfield  {journal}
  {\bibinfo  {journal} {Phys. Rev. Lett.}\ }\textbf {\bibinfo {volume} {101}},\
  \bibinfo {pages} {120406} (\bibinfo {year} {2008})}\BibitemShut {NoStop}%
\bibitem [{\citenamefont {Weller}\ \emph {et~al.}(2008)\citenamefont {Weller},
  \citenamefont {Ronzheimer}, \citenamefont {Gross}, \citenamefont {Esteve},
  \citenamefont {Oberthaler}, \citenamefont {Frantzeskakis}, \citenamefont
  {Theocharis},\ and\ \citenamefont {Kevrekidis}}]{PhysRevLett.101.130401}%
  \BibitemOpen
  \bibfield  {author} {\bibinfo {author} {\bibfnamefont {A.}~\bibnamefont
  {Weller}}, \bibinfo {author} {\bibfnamefont {J.~P.}\ \bibnamefont
  {Ronzheimer}}, \bibinfo {author} {\bibfnamefont {C.}~\bibnamefont {Gross}},
  \bibinfo {author} {\bibfnamefont {J.}~\bibnamefont {Esteve}}, \bibinfo
  {author} {\bibfnamefont {M.~K.}\ \bibnamefont {Oberthaler}}, \bibinfo
  {author} {\bibfnamefont {D.~J.}\ \bibnamefont {Frantzeskakis}}, \bibinfo
  {author} {\bibfnamefont {G.}~\bibnamefont {Theocharis}},\ and\ \bibinfo
  {author} {\bibfnamefont {P.~G.}\ \bibnamefont {Kevrekidis}},\ }\bibfield
  {title} {\bibinfo {title} {Experimental observation of oscillating and
  interacting matter wave dark solitons},\ }\href
  {https://doi.org/10.1103/PhysRevLett.101.130401} {\bibfield  {journal}
  {\bibinfo  {journal} {Phys. Rev. Lett.}\ }\textbf {\bibinfo {volume} {101}},\
  \bibinfo {pages} {130401} (\bibinfo {year} {2008})}\BibitemShut {NoStop}%
\bibitem [{\citenamefont {Chai}\ \emph {et~al.}(2020)\citenamefont {Chai},
  \citenamefont {Lao}, \citenamefont {Fujimoto}, \citenamefont {Hamazaki},
  \citenamefont {Ueda},\ and\ \citenamefont {Raman}}]{PhysRevLett.125.030402}%
  \BibitemOpen
  \bibfield  {author} {\bibinfo {author} {\bibfnamefont {X.}~\bibnamefont
  {Chai}}, \bibinfo {author} {\bibfnamefont {D.}~\bibnamefont {Lao}}, \bibinfo
  {author} {\bibfnamefont {K.}~\bibnamefont {Fujimoto}}, \bibinfo {author}
  {\bibfnamefont {R.}~\bibnamefont {Hamazaki}}, \bibinfo {author}
  {\bibfnamefont {M.}~\bibnamefont {Ueda}},\ and\ \bibinfo {author}
  {\bibfnamefont {C.}~\bibnamefont {Raman}},\ }\bibfield  {title} {\bibinfo
  {title} {Magnetic solitons in a spin-1 bose-einstein condensate},\ }\href
  {https://doi.org/10.1103/PhysRevLett.125.030402} {\bibfield  {journal}
  {\bibinfo  {journal} {Phys. Rev. Lett.}\ }\textbf {\bibinfo {volume} {125}},\
  \bibinfo {pages} {030402} (\bibinfo {year} {2020})}\BibitemShut {NoStop}%
\bibitem [{\citenamefont {Farolfi}\ \emph {et~al.}(2020)\citenamefont
  {Farolfi}, \citenamefont {Trypogeorgos}, \citenamefont {Mordini},
  \citenamefont {Lamporesi},\ and\ \citenamefont
  {Ferrari}}]{PhysRevLett.125.030401}%
  \BibitemOpen
  \bibfield  {author} {\bibinfo {author} {\bibfnamefont {A.}~\bibnamefont
  {Farolfi}}, \bibinfo {author} {\bibfnamefont {D.}~\bibnamefont
  {Trypogeorgos}}, \bibinfo {author} {\bibfnamefont {C.}~\bibnamefont
  {Mordini}}, \bibinfo {author} {\bibfnamefont {G.}~\bibnamefont {Lamporesi}},\
  and\ \bibinfo {author} {\bibfnamefont {G.}~\bibnamefont {Ferrari}},\
  }\bibfield  {title} {\bibinfo {title} {Observation of magnetic solitons in
  two-component bose-einstein condensates},\ }\href
  {https://doi.org/10.1103/PhysRevLett.125.030401} {\bibfield  {journal}
  {\bibinfo  {journal} {Phys. Rev. Lett.}\ }\textbf {\bibinfo {volume} {125}},\
  \bibinfo {pages} {030401} (\bibinfo {year} {2020})}\BibitemShut {NoStop}%
\bibitem [{\citenamefont {Aycock}\ \emph {et~al.}(2017)\citenamefont {Aycock},
  \citenamefont {Hurst}, \citenamefont {Efimkin}, \citenamefont {Genkina},
  \citenamefont {Lu}, \citenamefont {Galitski},\ and\ \citenamefont
  {Spielman}}]{PNAS.114.2503}%
  \BibitemOpen
  \bibfield  {author} {\bibinfo {author} {\bibfnamefont {L.~M.}\ \bibnamefont
  {Aycock}}, \bibinfo {author} {\bibfnamefont {H.~M.}\ \bibnamefont {Hurst}},
  \bibinfo {author} {\bibfnamefont {D.~K.}\ \bibnamefont {Efimkin}}, \bibinfo
  {author} {\bibfnamefont {D.}~\bibnamefont {Genkina}}, \bibinfo {author}
  {\bibfnamefont {H.-I.}\ \bibnamefont {Lu}}, \bibinfo {author} {\bibfnamefont
  {V.~M.}\ \bibnamefont {Galitski}},\ and\ \bibinfo {author} {\bibfnamefont
  {I.}~\bibnamefont {Spielman}},\ }\bibfield  {title} {\bibinfo {title}
  {Brownian motion of solitons in a bose--einstein condensate},\ }\href@noop {}
  {\bibfield  {journal} {\bibinfo  {journal} {PNAS}\ }\textbf {\bibinfo
  {volume} {114}},\ \bibinfo {pages} {2503} (\bibinfo {year}
  {2017})}\BibitemShut {NoStop}%
\bibitem [{\citenamefont {Meyer}\ \emph {et~al.}(2017)\citenamefont {Meyer},
  \citenamefont {Proud}, \citenamefont {Perea-Ortiz}, \citenamefont {O'Neale},
  \citenamefont {Baumert}, \citenamefont {Holynski}, \citenamefont
  {Kronj\"ager}, \citenamefont {Barontini},\ and\ \citenamefont
  {Bongs}}]{PhysRevLett.119.150403}%
  \BibitemOpen
  \bibfield  {author} {\bibinfo {author} {\bibfnamefont {N.}~\bibnamefont
  {Meyer}}, \bibinfo {author} {\bibfnamefont {H.}~\bibnamefont {Proud}},
  \bibinfo {author} {\bibfnamefont {M.}~\bibnamefont {Perea-Ortiz}}, \bibinfo
  {author} {\bibfnamefont {C.}~\bibnamefont {O'Neale}}, \bibinfo {author}
  {\bibfnamefont {M.}~\bibnamefont {Baumert}}, \bibinfo {author} {\bibfnamefont
  {M.}~\bibnamefont {Holynski}}, \bibinfo {author} {\bibfnamefont
  {J.}~\bibnamefont {Kronj\"ager}}, \bibinfo {author} {\bibfnamefont
  {G.}~\bibnamefont {Barontini}},\ and\ \bibinfo {author} {\bibfnamefont
  {K.}~\bibnamefont {Bongs}},\ }\bibfield  {title} {\bibinfo {title}
  {Observation of two-dimensional localized jones-roberts solitons in
  bose-einstein condensates},\ }\href
  {https://doi.org/10.1103/PhysRevLett.119.150403} {\bibfield  {journal}
  {\bibinfo  {journal} {Phys. Rev. Lett.}\ }\textbf {\bibinfo {volume} {119}},\
  \bibinfo {pages} {150403} (\bibinfo {year} {2017})}\BibitemShut {NoStop}%
\bibitem [{\citenamefont {Fritsch}\ \emph {et~al.}(2020)\citenamefont
  {Fritsch}, \citenamefont {Lu}, \citenamefont {Reid}, \citenamefont
  {Pi\~neiro},\ and\ \citenamefont {Spielman}}]{PhysRevA.101.053629}%
  \BibitemOpen
  \bibfield  {author} {\bibinfo {author} {\bibfnamefont {A.~R.}\ \bibnamefont
  {Fritsch}}, \bibinfo {author} {\bibfnamefont {M.}~\bibnamefont {Lu}},
  \bibinfo {author} {\bibfnamefont {G.~H.}\ \bibnamefont {Reid}}, \bibinfo
  {author} {\bibfnamefont {A.~M.}\ \bibnamefont {Pi\~neiro}},\ and\ \bibinfo
  {author} {\bibfnamefont {I.~B.}\ \bibnamefont {Spielman}},\ }\bibfield
  {title} {\bibinfo {title} {Creating solitons with controllable and near-zero
  velocity in bose-einstein condensates},\ }\href
  {https://doi.org/10.1103/PhysRevA.101.053629} {\bibfield  {journal} {\bibinfo
   {journal} {Phys. Rev. A}\ }\textbf {\bibinfo {volume} {101}},\ \bibinfo
  {pages} {053629} (\bibinfo {year} {2020})}\BibitemShut {NoStop}%
\bibitem [{\citenamefont {Dum}\ \emph {et~al.}(1998)\citenamefont {Dum},
  \citenamefont {Cirac}, \citenamefont {Lewenstein},\ and\ \citenamefont
  {Zoller}}]{PhysRevLett.80.2972}%
  \BibitemOpen
  \bibfield  {author} {\bibinfo {author} {\bibfnamefont {R.}~\bibnamefont
  {Dum}}, \bibinfo {author} {\bibfnamefont {J.~I.}\ \bibnamefont {Cirac}},
  \bibinfo {author} {\bibfnamefont {M.}~\bibnamefont {Lewenstein}},\ and\
  \bibinfo {author} {\bibfnamefont {P.}~\bibnamefont {Zoller}},\ }\bibfield
  {title} {\bibinfo {title} {Creation of dark solitons and vortices in
  bose-einstein condensates},\ }\href
  {https://doi.org/10.1103/PhysRevLett.80.2972} {\bibfield  {journal} {\bibinfo
   {journal} {Phys. Rev. Lett.}\ }\textbf {\bibinfo {volume} {80}},\ \bibinfo
  {pages} {2972} (\bibinfo {year} {1998})}\BibitemShut {NoStop}%
\bibitem [{\citenamefont {Carr}\ \emph {et~al.}(2001)\citenamefont {Carr},
  \citenamefont {Brand}, \citenamefont {Burger},\ and\ \citenamefont
  {Sanpera}}]{PhysRevA.63.051601}%
  \BibitemOpen
  \bibfield  {author} {\bibinfo {author} {\bibfnamefont {L.~D.}\ \bibnamefont
  {Carr}}, \bibinfo {author} {\bibfnamefont {J.}~\bibnamefont {Brand}},
  \bibinfo {author} {\bibfnamefont {S.}~\bibnamefont {Burger}},\ and\ \bibinfo
  {author} {\bibfnamefont {A.}~\bibnamefont {Sanpera}},\ }\bibfield  {title}
  {\bibinfo {title} {Dark-soliton creation in bose-einstein condensates},\
  }\href {https://doi.org/10.1103/PhysRevA.63.051601} {\bibfield  {journal}
  {\bibinfo  {journal} {Phys. Rev. A}\ }\textbf {\bibinfo {volume} {63}},\
  \bibinfo {pages} {051601} (\bibinfo {year} {2001})}\BibitemShut {NoStop}%
\bibitem [{\citenamefont {Scott}\ \emph {et~al.}(1998)\citenamefont {Scott},
  \citenamefont {Ballagh},\ and\ \citenamefont {Burnett}}]{Scott_1998}%
  \BibitemOpen
  \bibfield  {author} {\bibinfo {author} {\bibfnamefont {T.~F.}\ \bibnamefont
  {Scott}}, \bibinfo {author} {\bibfnamefont {R.~J.}\ \bibnamefont {Ballagh}},\
  and\ \bibinfo {author} {\bibfnamefont {K.}~\bibnamefont {Burnett}},\
  }\bibfield  {title} {\bibinfo {title} {Formation of fundamental structures in
  bose-einstein condensates},\ }\href@noop {} {\bibfield  {journal} {\bibinfo
  {journal} {J. Phys. B}\ }\textbf {\bibinfo {volume} {31}},\ \bibinfo {pages}
  {L329} (\bibinfo {year} {1998})}\BibitemShut {NoStop}%
\bibitem [{\citenamefont {Anderson}\ \emph {et~al.}(1995)\citenamefont
  {Anderson}, \citenamefont {Ensher}, \citenamefont {Matthews}, \citenamefont
  {Wieman},\ and\ \citenamefont {Cornell}}]{Anderson198}%
  \BibitemOpen
  \bibfield  {author} {\bibinfo {author} {\bibfnamefont {M.~H.}\ \bibnamefont
  {Anderson}}, \bibinfo {author} {\bibfnamefont {J.~R.}\ \bibnamefont
  {Ensher}}, \bibinfo {author} {\bibfnamefont {M.~R.}\ \bibnamefont
  {Matthews}}, \bibinfo {author} {\bibfnamefont {C.~E.}\ \bibnamefont
  {Wieman}},\ and\ \bibinfo {author} {\bibfnamefont {E.~A.}\ \bibnamefont
  {Cornell}},\ }\bibfield  {title} {\bibinfo {title} {Observation of
  bose-einstein condensation in a dilute atomic vapor},\ }\href@noop {}
  {\bibfield  {journal} {\bibinfo  {journal} {Science}\ }\textbf {\bibinfo
  {volume} {269}},\ \bibinfo {pages} {198} (\bibinfo {year}
  {1995})}\BibitemShut {NoStop}%
\bibitem [{\citenamefont {Davis}\ \emph {et~al.}(1995)\citenamefont {Davis},
  \citenamefont {Mewes}, \citenamefont {Andrews}, \citenamefont {van Druten},
  \citenamefont {Durfee}, \citenamefont {Kurn},\ and\ \citenamefont
  {Ketterle}}]{PhysRevLett.75.3969}%
  \BibitemOpen
  \bibfield  {author} {\bibinfo {author} {\bibfnamefont {K.~B.}\ \bibnamefont
  {Davis}}, \bibinfo {author} {\bibfnamefont {M.-O.}\ \bibnamefont {Mewes}},
  \bibinfo {author} {\bibfnamefont {M.~R.}\ \bibnamefont {Andrews}}, \bibinfo
  {author} {\bibfnamefont {N.~J.}\ \bibnamefont {van Druten}}, \bibinfo
  {author} {\bibfnamefont {D.~S.}\ \bibnamefont {Durfee}}, \bibinfo {author}
  {\bibfnamefont {D.}~\bibnamefont {Kurn}},\ and\ \bibinfo {author}
  {\bibfnamefont {W.}~\bibnamefont {Ketterle}},\ }\bibfield  {title} {\bibinfo
  {title} {Bose-einstein condensation in a gas of sodium atoms},\ }\href@noop
  {} {\bibfield  {journal} {\bibinfo  {journal} {Phys. Rev. Lett.}\ }\textbf
  {\bibinfo {volume} {75}},\ \bibinfo {pages} {3969} (\bibinfo {year}
  {1995})}\BibitemShut {NoStop}%
\bibitem [{\citenamefont {Bradley}\ \emph {et~al.}(1995)\citenamefont
  {Bradley}, \citenamefont {Sackett}, \citenamefont {Tollett},\ and\
  \citenamefont {Hulet}}]{PhysRevLett.75.1687}%
  \BibitemOpen
  \bibfield  {author} {\bibinfo {author} {\bibfnamefont {C.~C.}\ \bibnamefont
  {Bradley}}, \bibinfo {author} {\bibfnamefont {C.~A.}\ \bibnamefont
  {Sackett}}, \bibinfo {author} {\bibfnamefont {J.~J.}\ \bibnamefont
  {Tollett}},\ and\ \bibinfo {author} {\bibfnamefont {R.~G.}\ \bibnamefont
  {Hulet}},\ }\bibfield  {title} {\bibinfo {title} {Evidence of bose-einstein
  condensation in an atomic gas with attractive interactions},\ }\href
  {https://doi.org/10.1103/PhysRevLett.75.1687} {\bibfield  {journal} {\bibinfo
   {journal} {Phys. Rev. Lett.}\ }\textbf {\bibinfo {volume} {75}},\ \bibinfo
  {pages} {1687} (\bibinfo {year} {1995})}\BibitemShut {NoStop}%
\bibitem [{\citenamefont {Bradley}\ \emph
  {et~al.}(1997{\natexlab{a}})\citenamefont {Bradley}, \citenamefont {Sackett},
  \citenamefont {Tollett},\ and\ \citenamefont {Hulet}}]{PhysRevLett.79.1170}%
  \BibitemOpen
  \bibfield  {author} {\bibinfo {author} {\bibfnamefont {C.~C.}\ \bibnamefont
  {Bradley}}, \bibinfo {author} {\bibfnamefont {C.~A.}\ \bibnamefont
  {Sackett}}, \bibinfo {author} {\bibfnamefont {J.~J.}\ \bibnamefont
  {Tollett}},\ and\ \bibinfo {author} {\bibfnamefont {R.~G.}\ \bibnamefont
  {Hulet}},\ }\bibfield  {title} {\bibinfo {title} {Evidence of bose-einstein
  condensation in an atomic gas with attractive interactions [phys. rev. lett.
  75, 1687 (1995)]},\ }\href {https://doi.org/10.1103/PhysRevLett.79.1170}
  {\bibfield  {journal} {\bibinfo  {journal} {Phys. Rev. Lett.}\ }\textbf
  {\bibinfo {volume} {79}},\ \bibinfo {pages} {1170} (\bibinfo {year}
  {1997}{\natexlab{a}})}\BibitemShut {NoStop}%
\bibitem [{\citenamefont {G\"orlitz}\ \emph {et~al.}(2001)\citenamefont
  {G\"orlitz}, \citenamefont {Vogels}, \citenamefont {Leanhardt}, \citenamefont
  {Raman}, \citenamefont {Gustavson}, \citenamefont {Abo-Shaeer}, \citenamefont
  {Chikkatur}, \citenamefont {Gupta}, \citenamefont {Inouye}, \citenamefont
  {Rosenband},\ and\ \citenamefont {Ketterle}}]{PhysRevLett.87.130402}%
  \BibitemOpen
  \bibfield  {author} {\bibinfo {author} {\bibfnamefont {A.}~\bibnamefont
  {G\"orlitz}}, \bibinfo {author} {\bibfnamefont {J.~M.}\ \bibnamefont
  {Vogels}}, \bibinfo {author} {\bibfnamefont {A.~E.}\ \bibnamefont
  {Leanhardt}}, \bibinfo {author} {\bibfnamefont {C.}~\bibnamefont {Raman}},
  \bibinfo {author} {\bibfnamefont {T.~L.}\ \bibnamefont {Gustavson}}, \bibinfo
  {author} {\bibfnamefont {J.~R.}\ \bibnamefont {Abo-Shaeer}}, \bibinfo
  {author} {\bibfnamefont {A.~P.}\ \bibnamefont {Chikkatur}}, \bibinfo {author}
  {\bibfnamefont {S.}~\bibnamefont {Gupta}}, \bibinfo {author} {\bibfnamefont
  {S.}~\bibnamefont {Inouye}}, \bibinfo {author} {\bibfnamefont
  {T.}~\bibnamefont {Rosenband}},\ and\ \bibinfo {author} {\bibfnamefont
  {W.}~\bibnamefont {Ketterle}},\ }\bibfield  {title} {\bibinfo {title}
  {Realization of bose-einstein condensates in lower dimensions},\ }\href
  {https://doi.org/10.1103/PhysRevLett.87.130402} {\bibfield  {journal}
  {\bibinfo  {journal} {Phys. Rev. Lett.}\ }\textbf {\bibinfo {volume} {87}},\
  \bibinfo {pages} {130402} (\bibinfo {year} {2001})}\BibitemShut {NoStop}%
\bibitem [{\citenamefont {Andrelczyk}\ \emph {et~al.}(2001)\citenamefont
  {Andrelczyk}, \citenamefont {Brewczyk}, \citenamefont {Dobrek}, \citenamefont
  {Gajda},\ and\ \citenamefont {Lewenstein}}]{PhysRevA.64.043601}%
  \BibitemOpen
  \bibfield  {author} {\bibinfo {author} {\bibfnamefont {G.}~\bibnamefont
  {Andrelczyk}}, \bibinfo {author} {\bibfnamefont {M.}~\bibnamefont
  {Brewczyk}}, \bibinfo {author} {\bibfnamefont {L.}~\bibnamefont {Dobrek}},
  \bibinfo {author} {\bibfnamefont {M.}~\bibnamefont {Gajda}},\ and\ \bibinfo
  {author} {\bibfnamefont {M.}~\bibnamefont {Lewenstein}},\ }\bibfield  {title}
  {\bibinfo {title} {Optical generation of vortices in trapped bose-einstein
  condensates},\ }\href {https://doi.org/10.1103/PhysRevA.64.043601} {\bibfield
   {journal} {\bibinfo  {journal} {Phys. Rev. A}\ }\textbf {\bibinfo {volume}
  {64}},\ \bibinfo {pages} {043601} (\bibinfo {year} {2001})}\BibitemShut
  {NoStop}%
\bibitem [{\citenamefont {Romero-Ros}\ \emph {et~al.}(2020)\citenamefont
  {Romero-Ros}, \citenamefont {Katsimiga}, \citenamefont {Kevrekidis},
  \citenamefont {Prinari}, \citenamefont {Biondini},\ and\ \citenamefont
  {Schmelcher}}]{OnDemandPrePrint}%
  \BibitemOpen
  \bibfield  {author} {\bibinfo {author} {\bibfnamefont {A.}~\bibnamefont
  {Romero-Ros}}, \bibinfo {author} {\bibfnamefont {G.~C.}\ \bibnamefont
  {Katsimiga}}, \bibinfo {author} {\bibfnamefont {P.~G.}\ \bibnamefont
  {Kevrekidis}}, \bibinfo {author} {\bibfnamefont {B.}~\bibnamefont {Prinari}},
  \bibinfo {author} {\bibfnamefont {G.}~\bibnamefont {Biondini}},\ and\
  \bibinfo {author} {\bibfnamefont {P.}~\bibnamefont {Schmelcher}},\ }\bibfield
   {title} {\bibinfo {title} {On-demand generation of dark soliton trains in
  bose-einstein condensates},\ }\href@noop {} {\bibfield  {journal} {\bibinfo
  {journal} {arXiv preprint arXiv:2009.02292}\ } (\bibinfo {year}
  {2020})}\BibitemShut {NoStop}%
\bibitem [{\citenamefont {Halperin}\ and\ \citenamefont
  {Bohn}(2020)}]{halperin2020quench}%
  \BibitemOpen
  \bibfield  {author} {\bibinfo {author} {\bibfnamefont {E.~J.}\ \bibnamefont
  {Halperin}}\ and\ \bibinfo {author} {\bibfnamefont {J.~L.}\ \bibnamefont
  {Bohn}},\ }\bibfield  {title} {\bibinfo {title} {Quench-produced solitons in
  a box-trapped bose-einstein condensate},\ }\href@noop {} {\bibfield
  {journal} {\bibinfo  {journal} {arXiv preprint arXiv:2002.10491}\ } (\bibinfo
  {year} {2020})}\BibitemShut {NoStop}%
\bibitem [{\citenamefont {Kiehn}\ \emph {et~al.}(2019)\citenamefont {Kiehn},
  \citenamefont {Mistakidis}, \citenamefont {Katsimiga},\ and\ \citenamefont
  {Schmelcher}}]{PhysRevA.100.023613}%
  \BibitemOpen
  \bibfield  {author} {\bibinfo {author} {\bibfnamefont {H.}~\bibnamefont
  {Kiehn}}, \bibinfo {author} {\bibfnamefont {S.~I.}\ \bibnamefont
  {Mistakidis}}, \bibinfo {author} {\bibfnamefont {G.~C.}\ \bibnamefont
  {Katsimiga}},\ and\ \bibinfo {author} {\bibfnamefont {P.}~\bibnamefont
  {Schmelcher}},\ }\bibfield  {title} {\bibinfo {title} {Spontaneous generation
  of dark-bright and dark-antidark solitons upon quenching a
  particle-imbalanced bosonic mixture},\ }\href
  {https://doi.org/10.1103/PhysRevA.100.023613} {\bibfield  {journal} {\bibinfo
   {journal} {Phys. Rev. A}\ }\textbf {\bibinfo {volume} {100}},\ \bibinfo
  {pages} {023613} (\bibinfo {year} {2019})}\BibitemShut {NoStop}%
\bibitem [{\citenamefont {Theocharis}\ \emph {et~al.}(2007)\citenamefont
  {Theocharis}, \citenamefont {Kevrekidis}, \citenamefont {Oberthaler},\ and\
  \citenamefont {Frantzeskakis}}]{PhysRevA.76.045601}%
  \BibitemOpen
  \bibfield  {author} {\bibinfo {author} {\bibfnamefont {G.}~\bibnamefont
  {Theocharis}}, \bibinfo {author} {\bibfnamefont {P.~G.}\ \bibnamefont
  {Kevrekidis}}, \bibinfo {author} {\bibfnamefont {M.~K.}\ \bibnamefont
  {Oberthaler}},\ and\ \bibinfo {author} {\bibfnamefont {D.~J.}\ \bibnamefont
  {Frantzeskakis}},\ }\bibfield  {title} {\bibinfo {title} {Dark matter-wave
  solitons in the dimensionality crossover},\ }\href
  {https://doi.org/10.1103/PhysRevA.76.045601} {\bibfield  {journal} {\bibinfo
  {journal} {Phys. Rev. A}\ }\textbf {\bibinfo {volume} {76}},\ \bibinfo
  {pages} {045601} (\bibinfo {year} {2007})}\BibitemShut {NoStop}%
\bibitem [{\citenamefont {Muryshev}\ \emph {et~al.}(1999)\citenamefont
  {Muryshev}, \citenamefont {van Linden van~den Heuvell},\ and\ \citenamefont
  {Shlyapnikov}}]{PhysRevA.60.R2665}%
  \BibitemOpen
  \bibfield  {author} {\bibinfo {author} {\bibfnamefont {A.~E.}\ \bibnamefont
  {Muryshev}}, \bibinfo {author} {\bibfnamefont {H.~B.}\ \bibnamefont {van
  Linden van~den Heuvell}},\ and\ \bibinfo {author} {\bibfnamefont {G.~V.}\
  \bibnamefont {Shlyapnikov}},\ }\bibfield  {title} {\bibinfo {title}
  {Stability of standing matter waves in a trap},\ }\href
  {https://doi.org/10.1103/PhysRevA.60.R2665} {\bibfield  {journal} {\bibinfo
  {journal} {Phys. Rev. A}\ }\textbf {\bibinfo {volume} {60}},\ \bibinfo
  {pages} {R2665} (\bibinfo {year} {1999})}\BibitemShut {NoStop}%
\bibitem [{\citenamefont {Carr}\ \emph {et~al.}(2000)\citenamefont {Carr},
  \citenamefont {Clark},\ and\ \citenamefont {Reinhardt}}]{JPhysB.33.3983}%
  \BibitemOpen
  \bibfield  {author} {\bibinfo {author} {\bibfnamefont {L.~D.}\ \bibnamefont
  {Carr}}, \bibinfo {author} {\bibfnamefont {C.~W.}\ \bibnamefont {Clark}},\
  and\ \bibinfo {author} {\bibfnamefont {W.~P.}\ \bibnamefont {Reinhardt}},\
  }\bibfield  {title} {\bibinfo {title} {Stationary solutions of the
  one-dimensional nonlinear schr\"odinger equation. i. case of repulsive
  nonlinearity},\ }\href {https://doi.org/10.1103/PhysRevA.62.063610}
  {\bibfield  {journal} {\bibinfo  {journal} {Phys. Rev. A}\ }\textbf {\bibinfo
  {volume} {62}},\ \bibinfo {pages} {063610} (\bibinfo {year}
  {2000})}\BibitemShut {NoStop}%
\bibitem [{\citenamefont {Feder}\ \emph {et~al.}(2000)\citenamefont {Feder},
  \citenamefont {Pindzola}, \citenamefont {Collins}, \citenamefont
  {Schneider},\ and\ \citenamefont {Clark}}]{PhysRevA.62.053606}%
  \BibitemOpen
  \bibfield  {author} {\bibinfo {author} {\bibfnamefont {D.~L.}\ \bibnamefont
  {Feder}}, \bibinfo {author} {\bibfnamefont {M.~S.}\ \bibnamefont {Pindzola}},
  \bibinfo {author} {\bibfnamefont {L.~A.}\ \bibnamefont {Collins}}, \bibinfo
  {author} {\bibfnamefont {B.~I.}\ \bibnamefont {Schneider}},\ and\ \bibinfo
  {author} {\bibfnamefont {C.~W.}\ \bibnamefont {Clark}},\ }\bibfield  {title}
  {\bibinfo {title} {Dark-soliton states of bose-einstein condensates in
  anisotropic traps},\ }\href {https://doi.org/10.1103/PhysRevA.62.053606}
  {\bibfield  {journal} {\bibinfo  {journal} {Phys. Rev. A}\ }\textbf {\bibinfo
  {volume} {62}},\ \bibinfo {pages} {053606} (\bibinfo {year}
  {2000})}\BibitemShut {NoStop}%
\bibitem [{\citenamefont {Anderson}\ \emph {et~al.}(2001)\citenamefont
  {Anderson}, \citenamefont {Haljan}, \citenamefont {Regal}, \citenamefont
  {Feder}, \citenamefont {Collins}, \citenamefont {Clark},\ and\ \citenamefont
  {Cornell}}]{PhysRevLett.86.2926}%
  \BibitemOpen
  \bibfield  {author} {\bibinfo {author} {\bibfnamefont {B.~P.}\ \bibnamefont
  {Anderson}}, \bibinfo {author} {\bibfnamefont {P.~C.}\ \bibnamefont
  {Haljan}}, \bibinfo {author} {\bibfnamefont {C.~A.}\ \bibnamefont {Regal}},
  \bibinfo {author} {\bibfnamefont {D.~L.}\ \bibnamefont {Feder}}, \bibinfo
  {author} {\bibfnamefont {L.~A.}\ \bibnamefont {Collins}}, \bibinfo {author}
  {\bibfnamefont {C.~W.}\ \bibnamefont {Clark}},\ and\ \bibinfo {author}
  {\bibfnamefont {E.~A.}\ \bibnamefont {Cornell}},\ }\bibfield  {title}
  {\bibinfo {title} {Watching dark solitons decay into vortex rings in a
  bose-einstein condensate},\ }\href
  {https://doi.org/10.1103/PhysRevLett.86.2926} {\bibfield  {journal} {\bibinfo
   {journal} {Phys. Rev. Lett.}\ }\textbf {\bibinfo {volume} {86}},\ \bibinfo
  {pages} {2926} (\bibinfo {year} {2001})}\BibitemShut {NoStop}%
\bibitem [{\citenamefont {Hoefer}\ and\ \citenamefont
  {Ilan}(2016)}]{PhysRevA.94.013609}%
  \BibitemOpen
  \bibfield  {author} {\bibinfo {author} {\bibfnamefont {M.~A.}\ \bibnamefont
  {Hoefer}}\ and\ \bibinfo {author} {\bibfnamefont {B.}~\bibnamefont {Ilan}},\
  }\bibfield  {title} {\bibinfo {title} {Onset of transverse instabilities of
  confined dark solitons},\ }\href {https://doi.org/10.1103/PhysRevA.94.013609}
  {\bibfield  {journal} {\bibinfo  {journal} {Phys. Rev. A}\ }\textbf {\bibinfo
  {volume} {94}},\ \bibinfo {pages} {013609} (\bibinfo {year}
  {2016})}\BibitemShut {NoStop}%
\bibitem [{\citenamefont {Fedichev}\ \emph {et~al.}(1999)\citenamefont
  {Fedichev}, \citenamefont {Muryshev},\ and\ \citenamefont
  {Shlyapnikov}}]{PhysRevA.60.3220}%
  \BibitemOpen
  \bibfield  {author} {\bibinfo {author} {\bibfnamefont {P.~O.}\ \bibnamefont
  {Fedichev}}, \bibinfo {author} {\bibfnamefont {A.~E.}\ \bibnamefont
  {Muryshev}},\ and\ \bibinfo {author} {\bibfnamefont {G.~V.}\ \bibnamefont
  {Shlyapnikov}},\ }\bibfield  {title} {\bibinfo {title} {Dissipative dynamics
  of a kink state in a bose-condensed gas},\ }\href
  {https://doi.org/10.1103/PhysRevA.60.3220} {\bibfield  {journal} {\bibinfo
  {journal} {Phys. Rev. A}\ }\textbf {\bibinfo {volume} {60}},\ \bibinfo
  {pages} {3220} (\bibinfo {year} {1999})}\BibitemShut {NoStop}%
\bibitem [{\citenamefont {Muryshev}\ \emph {et~al.}(2002)\citenamefont
  {Muryshev}, \citenamefont {Shlyapnikov}, \citenamefont {Ertmer},
  \citenamefont {Sengstock},\ and\ \citenamefont
  {Lewenstein}}]{PhysRevLett.89.110401}%
  \BibitemOpen
  \bibfield  {author} {\bibinfo {author} {\bibfnamefont {A.}~\bibnamefont
  {Muryshev}}, \bibinfo {author} {\bibfnamefont {G.~V.}\ \bibnamefont
  {Shlyapnikov}}, \bibinfo {author} {\bibfnamefont {W.}~\bibnamefont {Ertmer}},
  \bibinfo {author} {\bibfnamefont {K.}~\bibnamefont {Sengstock}},\ and\
  \bibinfo {author} {\bibfnamefont {M.}~\bibnamefont {Lewenstein}},\ }\bibfield
   {title} {\bibinfo {title} {Dynamics of dark solitons in elongated
  bose-einstein condensates},\ }\href
  {https://doi.org/10.1103/PhysRevLett.89.110401} {\bibfield  {journal}
  {\bibinfo  {journal} {Phys. Rev. Lett.}\ }\textbf {\bibinfo {volume} {89}},\
  \bibinfo {pages} {110401} (\bibinfo {year} {2002})}\BibitemShut {NoStop}%
\bibitem [{\citenamefont {Jackson}\ \emph {et~al.}(2007)\citenamefont
  {Jackson}, \citenamefont {Proukakis},\ and\ \citenamefont
  {Barenghi}}]{PhysRevA.75.051601}%
  \BibitemOpen
  \bibfield  {author} {\bibinfo {author} {\bibfnamefont {B.}~\bibnamefont
  {Jackson}}, \bibinfo {author} {\bibfnamefont {N.~P.}\ \bibnamefont
  {Proukakis}},\ and\ \bibinfo {author} {\bibfnamefont {C.~F.}\ \bibnamefont
  {Barenghi}},\ }\bibfield  {title} {\bibinfo {title} {Dark-soliton dynamics in
  bose-einstein condensates at finite temperature},\ }\href
  {https://doi.org/10.1103/PhysRevA.75.051601} {\bibfield  {journal} {\bibinfo
  {journal} {Phys. Rev. A}\ }\textbf {\bibinfo {volume} {75}},\ \bibinfo
  {pages} {051601} (\bibinfo {year} {2007})}\BibitemShut {NoStop}%
\bibitem [{\citenamefont {Busch}\ and\ \citenamefont
  {Anglin}(2000)}]{PhysRevLett.84.2298}%
  \BibitemOpen
  \bibfield  {author} {\bibinfo {author} {\bibfnamefont {T.}~\bibnamefont
  {Busch}}\ and\ \bibinfo {author} {\bibfnamefont {J.~R.}\ \bibnamefont
  {Anglin}},\ }\bibfield  {title} {\bibinfo {title} {Motion of dark solitons in
  trapped bose-einstein condensates},\ }\href
  {https://doi.org/10.1103/PhysRevLett.84.2298} {\bibfield  {journal} {\bibinfo
   {journal} {Phys. Rev. Lett.}\ }\textbf {\bibinfo {volume} {84}},\ \bibinfo
  {pages} {2298} (\bibinfo {year} {2000})}\BibitemShut {NoStop}%
\bibitem [{\citenamefont {Parker}\ \emph {et~al.}(2003)\citenamefont {Parker},
  \citenamefont {Proukakis}, \citenamefont {Leadbeater},\ and\ \citenamefont
  {Adams}}]{PhysRevLett.90.220401}%
  \BibitemOpen
  \bibfield  {author} {\bibinfo {author} {\bibfnamefont {N.~G.}\ \bibnamefont
  {Parker}}, \bibinfo {author} {\bibfnamefont {N.~P.}\ \bibnamefont
  {Proukakis}}, \bibinfo {author} {\bibfnamefont {M.}~\bibnamefont
  {Leadbeater}},\ and\ \bibinfo {author} {\bibfnamefont {C.~S.}\ \bibnamefont
  {Adams}},\ }\bibfield  {title} {\bibinfo {title} {Soliton-sound interactions
  in quasi-one-dimensional bose-einstein condensates},\ }\href
  {https://doi.org/10.1103/PhysRevLett.90.220401} {\bibfield  {journal}
  {\bibinfo  {journal} {Phys. Rev. Lett.}\ }\textbf {\bibinfo {volume} {90}},\
  \bibinfo {pages} {220401} (\bibinfo {year} {2003})}\BibitemShut {NoStop}%
\bibitem [{\citenamefont {Parker}\ \emph {et~al.}(2010)\citenamefont {Parker},
  \citenamefont {Proukakis},\ and\ \citenamefont {Adams}}]{PhysRevA.81.033606}%
  \BibitemOpen
  \bibfield  {author} {\bibinfo {author} {\bibfnamefont {N.~G.}\ \bibnamefont
  {Parker}}, \bibinfo {author} {\bibfnamefont {N.~P.}\ \bibnamefont
  {Proukakis}},\ and\ \bibinfo {author} {\bibfnamefont {C.~S.}\ \bibnamefont
  {Adams}},\ }\bibfield  {title} {\bibinfo {title} {Dark soliton decay due to
  trap anharmonicity in atomic bose-einstein condensates},\ }\href
  {https://doi.org/10.1103/PhysRevA.81.033606} {\bibfield  {journal} {\bibinfo
  {journal} {Phys. Rev. A}\ }\textbf {\bibinfo {volume} {81}},\ \bibinfo
  {pages} {033606} (\bibinfo {year} {2010})}\BibitemShut {NoStop}%
\bibitem [{\citenamefont {Sciacca}\ \emph {et~al.}(2017)\citenamefont
  {Sciacca}, \citenamefont {Barenghi},\ and\ \citenamefont
  {Parker}}]{PhysRevA.95.013628}%
  \BibitemOpen
  \bibfield  {author} {\bibinfo {author} {\bibfnamefont {M.}~\bibnamefont
  {Sciacca}}, \bibinfo {author} {\bibfnamefont {C.~F.}\ \bibnamefont
  {Barenghi}},\ and\ \bibinfo {author} {\bibfnamefont {N.~G.}\ \bibnamefont
  {Parker}},\ }\bibfield  {title} {\bibinfo {title} {Matter-wave dark solitons
  in boxlike traps},\ }\href {https://doi.org/10.1103/PhysRevA.95.013628}
  {\bibfield  {journal} {\bibinfo  {journal} {Phys. Rev. A}\ }\textbf {\bibinfo
  {volume} {95}},\ \bibinfo {pages} {013628} (\bibinfo {year}
  {2017})}\BibitemShut {NoStop}%
\bibitem [{\citenamefont {Shomroni}\ \emph {et~al.}(2009)\citenamefont
  {Shomroni}, \citenamefont {Lahoud}, \citenamefont {Levy},\ and\ \citenamefont
  {Steinhauer}}]{shomroni2009evidence}%
  \BibitemOpen
  \bibfield  {author} {\bibinfo {author} {\bibfnamefont {I.}~\bibnamefont
  {Shomroni}}, \bibinfo {author} {\bibfnamefont {E.}~\bibnamefont {Lahoud}},
  \bibinfo {author} {\bibfnamefont {S.}~\bibnamefont {Levy}},\ and\ \bibinfo
  {author} {\bibfnamefont {J.}~\bibnamefont {Steinhauer}},\ }\bibfield  {title}
  {\bibinfo {title} {Evidence for an oscillating soliton/vortex ring by density
  engineering of a bose--einstein condensate},\ }\href@noop {} {\bibfield
  {journal} {\bibinfo  {journal} {Nat. Phys}\ }\textbf {\bibinfo {volume}
  {5}},\ \bibinfo {pages} {193} (\bibinfo {year} {2009})}\BibitemShut {NoStop}%
\bibitem [{\citenamefont {Kevrekidis}\ \emph {et~al.}(2019)\citenamefont
  {Kevrekidis}, \citenamefont {Wang}, \citenamefont {Theocharis}, \citenamefont
  {Frantzeskakis}, \citenamefont {Carretero-Gonz\'alez},\ and\ \citenamefont
  {Anderson}}]{PhysRevA.100.033607}%
  \BibitemOpen
  \bibfield  {author} {\bibinfo {author} {\bibfnamefont {P.~G.}\ \bibnamefont
  {Kevrekidis}}, \bibinfo {author} {\bibfnamefont {W.}~\bibnamefont {Wang}},
  \bibinfo {author} {\bibfnamefont {G.}~\bibnamefont {Theocharis}}, \bibinfo
  {author} {\bibfnamefont {D.~J.}\ \bibnamefont {Frantzeskakis}}, \bibinfo
  {author} {\bibfnamefont {R.}~\bibnamefont {Carretero-Gonz\'alez}},\ and\
  \bibinfo {author} {\bibfnamefont {B.~P.}\ \bibnamefont {Anderson}},\
  }\bibfield  {title} {\bibinfo {title} {Dynamics of interacting dark soliton
  stripes},\ }\href {https://doi.org/10.1103/PhysRevA.100.033607} {\bibfield
  {journal} {\bibinfo  {journal} {Phys. Rev. A}\ }\textbf {\bibinfo {volume}
  {100}},\ \bibinfo {pages} {033607} (\bibinfo {year} {2019})}\BibitemShut
  {NoStop}%
\bibitem [{\citenamefont {Theocharis}\ \emph {et~al.}(2010)\citenamefont
  {Theocharis}, \citenamefont {Weller}, \citenamefont {Ronzheimer},
  \citenamefont {Gross}, \citenamefont {Oberthaler}, \citenamefont
  {Kevrekidis},\ and\ \citenamefont {Frantzeskakis}}]{PhysRevA.81.063604}%
  \BibitemOpen
  \bibfield  {author} {\bibinfo {author} {\bibfnamefont {G.}~\bibnamefont
  {Theocharis}}, \bibinfo {author} {\bibfnamefont {A.}~\bibnamefont {Weller}},
  \bibinfo {author} {\bibfnamefont {J.~P.}\ \bibnamefont {Ronzheimer}},
  \bibinfo {author} {\bibfnamefont {C.}~\bibnamefont {Gross}}, \bibinfo
  {author} {\bibfnamefont {M.~K.}\ \bibnamefont {Oberthaler}}, \bibinfo
  {author} {\bibfnamefont {P.~G.}\ \bibnamefont {Kevrekidis}},\ and\ \bibinfo
  {author} {\bibfnamefont {D.~J.}\ \bibnamefont {Frantzeskakis}},\ }\bibfield
  {title} {\bibinfo {title} {Multiple atomic dark solitons in cigar-shaped
  bose-einstein condensates},\ }\href
  {https://doi.org/10.1103/PhysRevA.81.063604} {\bibfield  {journal} {\bibinfo
  {journal} {Phys. Rev. A}\ }\textbf {\bibinfo {volume} {81}},\ \bibinfo
  {pages} {063604} (\bibinfo {year} {2010})}\BibitemShut {NoStop}%
\bibitem [{\citenamefont {Yan}\ \emph {et~al.}(2011)\citenamefont {Yan},
  \citenamefont {Chang}, \citenamefont {Hamner}, \citenamefont {Kevrekidis},
  \citenamefont {Engels}, \citenamefont {Achilleos}, \citenamefont
  {Frantzeskakis}, \citenamefont {Carretero-Gonz\'alez},\ and\ \citenamefont
  {Schmelcher}}]{PhysRevA.84.053630}%
  \BibitemOpen
  \bibfield  {author} {\bibinfo {author} {\bibfnamefont {D.}~\bibnamefont
  {Yan}}, \bibinfo {author} {\bibfnamefont {J.~J.}\ \bibnamefont {Chang}},
  \bibinfo {author} {\bibfnamefont {C.}~\bibnamefont {Hamner}}, \bibinfo
  {author} {\bibfnamefont {P.~G.}\ \bibnamefont {Kevrekidis}}, \bibinfo
  {author} {\bibfnamefont {P.}~\bibnamefont {Engels}}, \bibinfo {author}
  {\bibfnamefont {V.}~\bibnamefont {Achilleos}}, \bibinfo {author}
  {\bibfnamefont {D.~J.}\ \bibnamefont {Frantzeskakis}}, \bibinfo {author}
  {\bibfnamefont {R.}~\bibnamefont {Carretero-Gonz\'alez}},\ and\ \bibinfo
  {author} {\bibfnamefont {P.}~\bibnamefont {Schmelcher}},\ }\bibfield  {title}
  {\bibinfo {title} {Multiple dark-bright solitons in atomic bose-einstein
  condensates},\ }\href {https://doi.org/10.1103/PhysRevA.84.053630} {\bibfield
   {journal} {\bibinfo  {journal} {Phys. Rev. A}\ }\textbf {\bibinfo {volume}
  {84}},\ \bibinfo {pages} {053630} (\bibinfo {year} {2011})}\BibitemShut
  {NoStop}%
\bibitem [{\citenamefont {Middelkamp}\ \emph {et~al.}(2011)\citenamefont
  {Middelkamp}, \citenamefont {Chang}, \citenamefont {Hamner}, \citenamefont
  {Carretero-Gonz{\'a}lez}, \citenamefont {Kevrekidis}, \citenamefont
  {Achilleos}, \citenamefont {Frantzeskakis}, \citenamefont {Schmelcher},\ and\
  \citenamefont {Engels}}]{middelkamp2011dynamics}%
  \BibitemOpen
  \bibfield  {author} {\bibinfo {author} {\bibfnamefont {S.}~\bibnamefont
  {Middelkamp}}, \bibinfo {author} {\bibfnamefont {J.}~\bibnamefont {Chang}},
  \bibinfo {author} {\bibfnamefont {C.}~\bibnamefont {Hamner}}, \bibinfo
  {author} {\bibfnamefont {R.}~\bibnamefont {Carretero-Gonz{\'a}lez}}, \bibinfo
  {author} {\bibfnamefont {P.~G.}\ \bibnamefont {Kevrekidis}}, \bibinfo
  {author} {\bibfnamefont {V.}~\bibnamefont {Achilleos}}, \bibinfo {author}
  {\bibfnamefont {D.~J.}\ \bibnamefont {Frantzeskakis}}, \bibinfo {author}
  {\bibfnamefont {P.}~\bibnamefont {Schmelcher}},\ and\ \bibinfo {author}
  {\bibfnamefont {P.}~\bibnamefont {Engels}},\ }\bibfield  {title} {\bibinfo
  {title} {Dynamics of dark--bright solitons in cigar-shaped bose--einstein
  condensates},\ }\href@noop {} {\bibfield  {journal} {\bibinfo  {journal}
  {Phys. Lett. A}\ }\textbf {\bibinfo {volume} {375}},\ \bibinfo {pages} {642}
  (\bibinfo {year} {2011})}\BibitemShut {NoStop}%
\bibitem [{\citenamefont {Romero-Ros}\ \emph {et~al.}(2019)\citenamefont
  {Romero-Ros}, \citenamefont {Katsimiga}, \citenamefont {Kevrekidis},\ and\
  \citenamefont {Schmelcher}}]{PhysRevA.100.013626}%
  \BibitemOpen
  \bibfield  {author} {\bibinfo {author} {\bibfnamefont {A.}~\bibnamefont
  {Romero-Ros}}, \bibinfo {author} {\bibfnamefont {G.~C.}\ \bibnamefont
  {Katsimiga}}, \bibinfo {author} {\bibfnamefont {P.~G.}\ \bibnamefont
  {Kevrekidis}},\ and\ \bibinfo {author} {\bibfnamefont {P.}~\bibnamefont
  {Schmelcher}},\ }\bibfield  {title} {\bibinfo {title} {Controlled generation
  of dark-bright soliton complexes in two-component and spinor bose-einstein
  condensates},\ }\href {https://doi.org/10.1103/PhysRevA.100.013626}
  {\bibfield  {journal} {\bibinfo  {journal} {Phys. Rev. A}\ }\textbf {\bibinfo
  {volume} {100}},\ \bibinfo {pages} {013626} (\bibinfo {year}
  {2019})}\BibitemShut {NoStop}%
\bibitem [{\citenamefont {Barenghi}\ \emph {et~al.}(2014)\citenamefont
  {Barenghi}, \citenamefont {Skrbek},\ and\ \citenamefont
  {Sreenivasan}}]{barenghi2014introduction}%
  \BibitemOpen
  \bibfield  {author} {\bibinfo {author} {\bibfnamefont {C.~F.}\ \bibnamefont
  {Barenghi}}, \bibinfo {author} {\bibfnamefont {L.}~\bibnamefont {Skrbek}},\
  and\ \bibinfo {author} {\bibfnamefont {K.~R.}\ \bibnamefont {Sreenivasan}},\
  }\bibfield  {title} {\bibinfo {title} {Introduction to quantum turbulence},\
  }\href@noop {} {\bibfield  {journal} {\bibinfo  {journal} {Proc. Natl. Acad.
  Sci. U.S.A}\ }\textbf {\bibinfo {volume} {111}},\ \bibinfo {pages} {4647}
  (\bibinfo {year} {2014})}\BibitemShut {NoStop}%
\bibitem [{\citenamefont {Tsatsos}\ \emph {et~al.}(2016)\citenamefont
  {Tsatsos}, \citenamefont {Tavares}, \citenamefont {Cidrim}, \citenamefont
  {Fritsch}, \citenamefont {Caracanhas}, \citenamefont {dos Santos},
  \citenamefont {Barenghi},\ and\ \citenamefont
  {Bagnato}}]{tsatsos2016quantum}%
  \BibitemOpen
  \bibfield  {author} {\bibinfo {author} {\bibfnamefont {M.~C.}\ \bibnamefont
  {Tsatsos}}, \bibinfo {author} {\bibfnamefont {P.~E.~S.}\ \bibnamefont
  {Tavares}}, \bibinfo {author} {\bibfnamefont {A.}~\bibnamefont {Cidrim}},
  \bibinfo {author} {\bibfnamefont {A.~R.}\ \bibnamefont {Fritsch}}, \bibinfo
  {author} {\bibfnamefont {M.~A.}\ \bibnamefont {Caracanhas}}, \bibinfo
  {author} {\bibfnamefont {F.~E.~A.}\ \bibnamefont {dos Santos}}, \bibinfo
  {author} {\bibfnamefont {C.~F.}\ \bibnamefont {Barenghi}},\ and\ \bibinfo
  {author} {\bibfnamefont {V.~S.}\ \bibnamefont {Bagnato}},\ }\bibfield
  {title} {\bibinfo {title} {Quantum turbulence in trapped atomic
  bose--einstein condensates},\ }\href@noop {} {\bibfield  {journal} {\bibinfo
  {journal} {Phys. Rep.}\ }\textbf {\bibinfo {volume} {622}},\ \bibinfo {pages}
  {1} (\bibinfo {year} {2016})}\BibitemShut {NoStop}%
\bibitem [{\citenamefont {Henn}\ \emph {et~al.}(2009)\citenamefont {Henn},
  \citenamefont {Seman}, \citenamefont {Roati}, \citenamefont {Magalh\~aes},\
  and\ \citenamefont {Bagnato}}]{PhysRevLett.103.045301}%
  \BibitemOpen
  \bibfield  {author} {\bibinfo {author} {\bibfnamefont {E.~A.~L.}\
  \bibnamefont {Henn}}, \bibinfo {author} {\bibfnamefont {J.~A.}\ \bibnamefont
  {Seman}}, \bibinfo {author} {\bibfnamefont {G.}~\bibnamefont {Roati}},
  \bibinfo {author} {\bibfnamefont {K.~M.~F.}\ \bibnamefont {Magalh\~aes}},\
  and\ \bibinfo {author} {\bibfnamefont {V.~S.}\ \bibnamefont {Bagnato}},\
  }\bibfield  {title} {\bibinfo {title} {Emergence of turbulence in an
  oscillating bose-einstein condensate},\ }\href
  {https://doi.org/10.1103/PhysRevLett.103.045301} {\bibfield  {journal}
  {\bibinfo  {journal} {Phys. Rev. Lett.}\ }\textbf {\bibinfo {volume} {103}},\
  \bibinfo {pages} {045301} (\bibinfo {year} {2009})}\BibitemShut {NoStop}%
\bibitem [{\citenamefont {Navon}\ \emph {et~al.}(2016)\citenamefont {Navon},
  \citenamefont {Gaunt}, \citenamefont {Smith},\ and\ \citenamefont
  {Hadzibabic}}]{Nature.539.72}%
  \BibitemOpen
  \bibfield  {author} {\bibinfo {author} {\bibfnamefont {N.}~\bibnamefont
  {Navon}}, \bibinfo {author} {\bibfnamefont {A.~L.}\ \bibnamefont {Gaunt}},
  \bibinfo {author} {\bibfnamefont {R.~P.}\ \bibnamefont {Smith}},\ and\
  \bibinfo {author} {\bibfnamefont {Z.}~\bibnamefont {Hadzibabic}},\ }\bibfield
   {title} {\bibinfo {title} {Emergence of a turbulent cascade in a quantum
  gas},\ }\href@noop {} {\bibfield  {journal} {\bibinfo  {journal} {Nature}\
  }\textbf {\bibinfo {volume} {539}},\ \bibinfo {pages} {72} (\bibinfo {year}
  {2016})}\BibitemShut {NoStop}%
\bibitem [{\citenamefont {Andrews}\ \emph {et~al.}(1996)\citenamefont
  {Andrews}, \citenamefont {Mewes}, \citenamefont {Van~Druten}, \citenamefont
  {Durfee}, \citenamefont {Kurn},\ and\ \citenamefont {Ketterle}}]{Andrews84}%
  \BibitemOpen
  \bibfield  {author} {\bibinfo {author} {\bibfnamefont {M.~R.}\ \bibnamefont
  {Andrews}}, \bibinfo {author} {\bibfnamefont {M.-O.}\ \bibnamefont {Mewes}},
  \bibinfo {author} {\bibfnamefont {N.~J.}\ \bibnamefont {Van~Druten}},
  \bibinfo {author} {\bibfnamefont {D.~S.}\ \bibnamefont {Durfee}}, \bibinfo
  {author} {\bibfnamefont {D.~M.}\ \bibnamefont {Kurn}},\ and\ \bibinfo
  {author} {\bibfnamefont {W.}~\bibnamefont {Ketterle}},\ }\bibfield  {title}
  {\bibinfo {title} {Direct, nondestructive observation of a bose condensate},\
  }\href@noop {} {\bibfield  {journal} {\bibinfo  {journal} {Science}\ }\textbf
  {\bibinfo {volume} {273}},\ \bibinfo {pages} {84} (\bibinfo {year}
  {1996})}\BibitemShut {NoStop}%
\bibitem [{\citenamefont {Ramanathan}\ \emph
  {et~al.}(2012{\natexlab{a}})\citenamefont {Ramanathan}, \citenamefont
  {Muniz}, \citenamefont {Wright}, \citenamefont {Anderson}, \citenamefont
  {Phillips}, \citenamefont {Helmerson},\ and\ \citenamefont
  {Campbell}}]{doi:10.1063/1.4747163}%
  \BibitemOpen
  \bibfield  {author} {\bibinfo {author} {\bibfnamefont {A.}~\bibnamefont
  {Ramanathan}}, \bibinfo {author} {\bibfnamefont {S.~R.}\ \bibnamefont
  {Muniz}}, \bibinfo {author} {\bibfnamefont {K.~C.}\ \bibnamefont {Wright}},
  \bibinfo {author} {\bibfnamefont {R.~P.}\ \bibnamefont {Anderson}}, \bibinfo
  {author} {\bibfnamefont {W.~D.}\ \bibnamefont {Phillips}}, \bibinfo {author}
  {\bibfnamefont {K.}~\bibnamefont {Helmerson}},\ and\ \bibinfo {author}
  {\bibfnamefont {G.~K.}\ \bibnamefont {Campbell}},\ }\bibfield  {title}
  {\bibinfo {title} {Partial-transfer absorption imaging: A versatile technique
  for optimal imaging of ultracold gases},\ }\href@noop {} {\bibfield
  {journal} {\bibinfo  {journal} {Rev. Sci. Instrum.}\ }\textbf {\bibinfo
  {volume} {83}},\ \bibinfo {pages} {083119} (\bibinfo {year}
  {2012}{\natexlab{a}})}\BibitemShut {NoStop}%
\bibitem [{\citenamefont {Streed}\ \emph {et~al.}(2012)\citenamefont {Streed},
  \citenamefont {Jechow}, \citenamefont {Norton},\ and\ \citenamefont
  {Kielpinski}}]{absorption}%
  \BibitemOpen
  \bibfield  {author} {\bibinfo {author} {\bibfnamefont {E.~W.}\ \bibnamefont
  {Streed}}, \bibinfo {author} {\bibfnamefont {A.}~\bibnamefont {Jechow}},
  \bibinfo {author} {\bibfnamefont {B.~G.}\ \bibnamefont {Norton}},\ and\
  \bibinfo {author} {\bibfnamefont {D.}~\bibnamefont {Kielpinski}},\ }\bibfield
   {title} {\bibinfo {title} {Absorption imaging of a single atom},\
  }\href@noop {} {\bibfield  {journal} {\bibinfo  {journal} {Nature
  communications}\ }\textbf {\bibinfo {volume} {3}},\ \bibinfo {pages} {1}
  (\bibinfo {year} {2012})}\BibitemShut {NoStop}%
\bibitem [{\citenamefont {Bradley}\ \emph
  {et~al.}(1997{\natexlab{b}})\citenamefont {Bradley}, \citenamefont
  {Sackett},\ and\ \citenamefont {Hulet}}]{PhysRevLett.78.985}%
  \BibitemOpen
  \bibfield  {author} {\bibinfo {author} {\bibfnamefont {C.~C.}\ \bibnamefont
  {Bradley}}, \bibinfo {author} {\bibfnamefont {C.~A.}\ \bibnamefont
  {Sackett}},\ and\ \bibinfo {author} {\bibfnamefont {R.~G.}\ \bibnamefont
  {Hulet}},\ }\bibfield  {title} {\bibinfo {title} {Bose-einstein condensation
  of lithium: Observation of limited condensate number},\ }\href
  {https://doi.org/10.1103/PhysRevLett.78.985} {\bibfield  {journal} {\bibinfo
  {journal} {Phys. Rev. Lett.}\ }\textbf {\bibinfo {volume} {78}},\ \bibinfo
  {pages} {985} (\bibinfo {year} {1997}{\natexlab{b}})}\BibitemShut {NoStop}%
\bibitem [{\citenamefont {Meppelink}\ \emph {et~al.}(2010)\citenamefont
  {Meppelink}, \citenamefont {Rozendaal}, \citenamefont {Koller}, \citenamefont
  {Vogels},\ and\ \citenamefont {Van~der Straten}}]{PhysRevA.81.053632}%
  \BibitemOpen
  \bibfield  {author} {\bibinfo {author} {\bibfnamefont {R.}~\bibnamefont
  {Meppelink}}, \bibinfo {author} {\bibfnamefont {R.~A.}\ \bibnamefont
  {Rozendaal}}, \bibinfo {author} {\bibfnamefont {S.~B.}\ \bibnamefont
  {Koller}}, \bibinfo {author} {\bibfnamefont {J.~M.}\ \bibnamefont {Vogels}},\
  and\ \bibinfo {author} {\bibfnamefont {P.}~\bibnamefont {Van~der Straten}},\
  }\bibfield  {title} {\bibinfo {title} {Thermodynamics of
  bose-einstein-condensed clouds using phase-contrast imaging},\ }\href@noop {}
  {\bibfield  {journal} {\bibinfo  {journal} {Physical Review A}\ }\textbf
  {\bibinfo {volume} {81}},\ \bibinfo {pages} {053632} (\bibinfo {year}
  {2010})}\BibitemShut {NoStop}%
\bibitem [{\citenamefont {Hope}\ and\ \citenamefont
  {Close}(2004)}]{PhysRevLett.93.180402}%
  \BibitemOpen
  \bibfield  {author} {\bibinfo {author} {\bibfnamefont {J.~J.}\ \bibnamefont
  {Hope}}\ and\ \bibinfo {author} {\bibfnamefont {J.~D.}\ \bibnamefont
  {Close}},\ }\bibfield  {title} {\bibinfo {title} {Limit to minimally
  destructive optical detection of atoms},\ }\href
  {https://doi.org/10.1103/PhysRevLett.93.180402} {\bibfield  {journal}
  {\bibinfo  {journal} {Phys. Rev. Lett.}\ }\textbf {\bibinfo {volume} {93}},\
  \bibinfo {pages} {180402} (\bibinfo {year} {2004})}\BibitemShut {NoStop}%
\bibitem [{\citenamefont {Gauthier}\ \emph {et~al.}(2016)\citenamefont
  {Gauthier}, \citenamefont {Lenton}, \citenamefont {Parry}, \citenamefont
  {Baker}, \citenamefont {Davis}, \citenamefont {Rubinsztein-Dunlop},\ and\
  \citenamefont {Neely}}]{Gauthier:s}%
  \BibitemOpen
  \bibfield  {author} {\bibinfo {author} {\bibfnamefont {G.}~\bibnamefont
  {Gauthier}}, \bibinfo {author} {\bibfnamefont {I.}~\bibnamefont {Lenton}},
  \bibinfo {author} {\bibfnamefont {N.~M.}\ \bibnamefont {Parry}}, \bibinfo
  {author} {\bibfnamefont {M.}~\bibnamefont {Baker}}, \bibinfo {author}
  {\bibfnamefont {M.}~\bibnamefont {Davis}}, \bibinfo {author} {\bibfnamefont
  {H.}~\bibnamefont {Rubinsztein-Dunlop}},\ and\ \bibinfo {author}
  {\bibfnamefont {T.}~\bibnamefont {Neely}},\ }\bibfield  {title} {\bibinfo
  {title} {Direct imaging of a digital-micromirror device for configurable
  microscopic optical potentials},\ }\href@noop {} {\bibfield  {journal}
  {\bibinfo  {journal} {Optica}\ }\textbf {\bibinfo {volume} {3}},\ \bibinfo
  {pages} {1136} (\bibinfo {year} {2016})}\BibitemShut {NoStop}%
\bibitem [{\citenamefont {Wilson}\ \emph {et~al.}(2015)\citenamefont {Wilson},
  \citenamefont {Newman}, \citenamefont {Lowney},\ and\ \citenamefont
  {Anderson}}]{PhysRevA.91.023621}%
  \BibitemOpen
  \bibfield  {author} {\bibinfo {author} {\bibfnamefont {K.~E.}\ \bibnamefont
  {Wilson}}, \bibinfo {author} {\bibfnamefont {Z.~L.}\ \bibnamefont {Newman}},
  \bibinfo {author} {\bibfnamefont {J.~D.}\ \bibnamefont {Lowney}},\ and\
  \bibinfo {author} {\bibfnamefont {B.~P.}\ \bibnamefont {Anderson}},\
  }\bibfield  {title} {\bibinfo {title} {In situ imaging of vortices in
  bose-einstein condensates},\ }\href
  {https://doi.org/10.1103/PhysRevA.91.023621} {\bibfield  {journal} {\bibinfo
  {journal} {Phys. Rev. A}\ }\textbf {\bibinfo {volume} {91}},\ \bibinfo
  {pages} {023621} (\bibinfo {year} {2015})}\BibitemShut {NoStop}%
\bibitem [{\citenamefont {Seo}\ \emph {et~al.}(2017)\citenamefont {Seo},
  \citenamefont {Ko}, \citenamefont {Kim},\ and\ \citenamefont
  {Shin}}]{Shin_2D}%
  \BibitemOpen
  \bibfield  {author} {\bibinfo {author} {\bibfnamefont {S.~W.}\ \bibnamefont
  {Seo}}, \bibinfo {author} {\bibfnamefont {B.}~\bibnamefont {Ko}}, \bibinfo
  {author} {\bibfnamefont {J.~H.}\ \bibnamefont {Kim}},\ and\ \bibinfo {author}
  {\bibfnamefont {Y.-i.}\ \bibnamefont {Shin}},\ }\bibfield  {title} {\bibinfo
  {title} {Observation of vortex-antivortex pairing in decaying 2d turbulence
  of a superfluid gas},\ }\href@noop {} {\bibfield  {journal} {\bibinfo
  {journal} {Sci. Rep.}\ }\textbf {\bibinfo {volume} {7}},\ \bibinfo {pages}
  {1} (\bibinfo {year} {2017})}\BibitemShut {NoStop}%
\bibitem [{\citenamefont {Freilich}\ \emph {et~al.}(2010)\citenamefont
  {Freilich}, \citenamefont {Bianchi}, \citenamefont {Kaufman}, \citenamefont
  {Langin},\ and\ \citenamefont {Hall}}]{Freilich1182}%
  \BibitemOpen
  \bibfield  {author} {\bibinfo {author} {\bibfnamefont {D.~V.}\ \bibnamefont
  {Freilich}}, \bibinfo {author} {\bibfnamefont {D.~M.}\ \bibnamefont
  {Bianchi}}, \bibinfo {author} {\bibfnamefont {A.~M.}\ \bibnamefont
  {Kaufman}}, \bibinfo {author} {\bibfnamefont {T.~K.}\ \bibnamefont
  {Langin}},\ and\ \bibinfo {author} {\bibfnamefont {D.~S.}\ \bibnamefont
  {Hall}},\ }\bibfield  {title} {\bibinfo {title} {Real-time dynamics of single
  vortex lines and vortex dipoles in a bose-einstein condensate},\ }\href@noop
  {} {\bibfield  {journal} {\bibinfo  {journal} {Science}\ }\textbf {\bibinfo
  {volume} {329}},\ \bibinfo {pages} {1182} (\bibinfo {year}
  {2010})}\BibitemShut {NoStop}%
\bibitem [{\citenamefont {Ramanathan}\ \emph
  {et~al.}(2012{\natexlab{b}})\citenamefont {Ramanathan}, \citenamefont
  {Muniz}, \citenamefont {Wright}, \citenamefont {Anderson}, \citenamefont
  {Phillips}, \citenamefont {Helmerson},\ and\ \citenamefont
  {Campbell}}]{Seroka:19}%
  \BibitemOpen
  \bibfield  {author} {\bibinfo {author} {\bibfnamefont {A.}~\bibnamefont
  {Ramanathan}}, \bibinfo {author} {\bibfnamefont {S.~R.}\ \bibnamefont
  {Muniz}}, \bibinfo {author} {\bibfnamefont {K.~C.}\ \bibnamefont {Wright}},
  \bibinfo {author} {\bibfnamefont {R.~P.}\ \bibnamefont {Anderson}}, \bibinfo
  {author} {\bibfnamefont {W.~D.}\ \bibnamefont {Phillips}}, \bibinfo {author}
  {\bibfnamefont {K.}~\bibnamefont {Helmerson}},\ and\ \bibinfo {author}
  {\bibfnamefont {G.~K.}\ \bibnamefont {Campbell}},\ }\bibfield  {title}
  {\bibinfo {title} {Partial-transfer absorption imaging: A versatile technique
  for optimal imaging of ultracold gases},\ }\href@noop {} {\bibfield
  {journal} {\bibinfo  {journal} {Rev. Sci. Instrum.}\ }\textbf {\bibinfo
  {volume} {83}},\ \bibinfo {pages} {083119} (\bibinfo {year}
  {2012}{\natexlab{b}})}\BibitemShut {NoStop}%
\bibitem [{\citenamefont {Donadello}\ \emph {et~al.}(2014)\citenamefont
  {Donadello}, \citenamefont {Serafini}, \citenamefont {Tylutki}, \citenamefont
  {Pitaevskii}, \citenamefont {Dalfovo}, \citenamefont {Lamporesi},\ and\
  \citenamefont {Ferrari}}]{Donadello2014}%
  \BibitemOpen
  \bibfield  {author} {\bibinfo {author} {\bibfnamefont {S.}~\bibnamefont
  {Donadello}}, \bibinfo {author} {\bibfnamefont {S.}~\bibnamefont {Serafini}},
  \bibinfo {author} {\bibfnamefont {M.}~\bibnamefont {Tylutki}}, \bibinfo
  {author} {\bibfnamefont {L.~P.}\ \bibnamefont {Pitaevskii}}, \bibinfo
  {author} {\bibfnamefont {F.}~\bibnamefont {Dalfovo}}, \bibinfo {author}
  {\bibfnamefont {G.}~\bibnamefont {Lamporesi}},\ and\ \bibinfo {author}
  {\bibfnamefont {G.}~\bibnamefont {Ferrari}},\ }\bibfield  {title} {\bibinfo
  {title} {Observation of solitonic vortices in bose-einstein condensates},\
  }\href {https://doi.org/10.1103/PhysRevLett.113.065302} {\bibfield  {journal}
  {\bibinfo  {journal} {Phys. Rev. Lett.}\ }\textbf {\bibinfo {volume} {113}},\
  \bibinfo {pages} {065302} (\bibinfo {year} {2014})}\BibitemShut {NoStop}%
\bibitem [{\citenamefont {Serafini}\ \emph {et~al.}(2015)\citenamefont
  {Serafini}, \citenamefont {Barbiero}, \citenamefont {Debortoli},
  \citenamefont {Donadello}, \citenamefont {Larcher}, \citenamefont {Dalfovo},
  \citenamefont {Lamporesi},\ and\ \citenamefont
  {Ferrari}}]{PhysRevLett.115.170402}%
  \BibitemOpen
  \bibfield  {author} {\bibinfo {author} {\bibfnamefont {S.}~\bibnamefont
  {Serafini}}, \bibinfo {author} {\bibfnamefont {M.}~\bibnamefont {Barbiero}},
  \bibinfo {author} {\bibfnamefont {M.}~\bibnamefont {Debortoli}}, \bibinfo
  {author} {\bibfnamefont {S.}~\bibnamefont {Donadello}}, \bibinfo {author}
  {\bibfnamefont {F.}~\bibnamefont {Larcher}}, \bibinfo {author} {\bibfnamefont
  {F.}~\bibnamefont {Dalfovo}}, \bibinfo {author} {\bibfnamefont
  {G.}~\bibnamefont {Lamporesi}},\ and\ \bibinfo {author} {\bibfnamefont
  {G.}~\bibnamefont {Ferrari}},\ }\bibfield  {title} {\bibinfo {title}
  {Dynamics and interaction of vortex lines in an elongated bose-einstein
  condensate},\ }\href {https://doi.org/10.1103/PhysRevLett.115.170402}
  {\bibfield  {journal} {\bibinfo  {journal} {Phys. Rev. Lett.}\ }\textbf
  {\bibinfo {volume} {115}},\ \bibinfo {pages} {170402} (\bibinfo {year}
  {2015})}\BibitemShut {NoStop}%
\bibitem [{\citenamefont {Serafini}\ \emph {et~al.}(2017)\citenamefont
  {Serafini}, \citenamefont {Galantucci}, \citenamefont {Iseni}, \citenamefont
  {Bienaim\'e}, \citenamefont {Bisset}, \citenamefont {Barenghi}, \citenamefont
  {Dalfovo}, \citenamefont {Lamporesi},\ and\ \citenamefont
  {Ferrari}}]{Serafini2017}%
  \BibitemOpen
  \bibfield  {author} {\bibinfo {author} {\bibfnamefont {S.}~\bibnamefont
  {Serafini}}, \bibinfo {author} {\bibfnamefont {L.}~\bibnamefont
  {Galantucci}}, \bibinfo {author} {\bibfnamefont {E.}~\bibnamefont {Iseni}},
  \bibinfo {author} {\bibfnamefont {T.}~\bibnamefont {Bienaim\'e}}, \bibinfo
  {author} {\bibfnamefont {R.~N.}\ \bibnamefont {Bisset}}, \bibinfo {author}
  {\bibfnamefont {C.~F.}\ \bibnamefont {Barenghi}}, \bibinfo {author}
  {\bibfnamefont {F.}~\bibnamefont {Dalfovo}}, \bibinfo {author} {\bibfnamefont
  {G.}~\bibnamefont {Lamporesi}},\ and\ \bibinfo {author} {\bibfnamefont
  {G.}~\bibnamefont {Ferrari}},\ }\bibfield  {title} {\bibinfo {title} {Vortex
  reconnections and rebounds in trapped atomic bose-einstein condensates},\
  }\href {https://doi.org/10.1103/PhysRevX.7.021031} {\bibfield  {journal}
  {\bibinfo  {journal} {Phys. Rev. X}\ }\textbf {\bibinfo {volume} {7}},\
  \bibinfo {pages} {021031} (\bibinfo {year} {2017})}\BibitemShut {NoStop}%
\bibitem [{\citenamefont {Hueck}\ \emph {et~al.}(2017)\citenamefont {Hueck},
  \citenamefont {Luick}, \citenamefont {Sobirey}, \citenamefont {Siegl},
  \citenamefont {Lompe}, \citenamefont {Moritz}, \citenamefont {Clark},\ and\
  \citenamefont {Chin}}]{Hueck:17}%
  \BibitemOpen
  \bibfield  {author} {\bibinfo {author} {\bibfnamefont {K.}~\bibnamefont
  {Hueck}}, \bibinfo {author} {\bibfnamefont {N.}~\bibnamefont {Luick}},
  \bibinfo {author} {\bibfnamefont {L.}~\bibnamefont {Sobirey}}, \bibinfo
  {author} {\bibfnamefont {J.}~\bibnamefont {Siegl}}, \bibinfo {author}
  {\bibfnamefont {T.}~\bibnamefont {Lompe}}, \bibinfo {author} {\bibfnamefont
  {H.}~\bibnamefont {Moritz}}, \bibinfo {author} {\bibfnamefont {L.~W.}\
  \bibnamefont {Clark}},\ and\ \bibinfo {author} {\bibfnamefont
  {C.}~\bibnamefont {Chin}},\ }\bibfield  {title} {\bibinfo {title}
  {Calibrating high intensity absorption imaging of ultracold atoms},\
  }\href@noop {} {\bibfield  {journal} {\bibinfo  {journal} {Opt. Express}\
  }\textbf {\bibinfo {volume} {25}},\ \bibinfo {pages} {8670} (\bibinfo {year}
  {2017})}\BibitemShut {NoStop}%
\bibitem [{\citenamefont {Putra}\ \emph {et~al.}(2014)\citenamefont {Putra},
  \citenamefont {Campbell}, \citenamefont {Price}, \citenamefont {De},\ and\
  \citenamefont {Spielman}}]{Putra}%
  \BibitemOpen
  \bibfield  {author} {\bibinfo {author} {\bibfnamefont {A.}~\bibnamefont
  {Putra}}, \bibinfo {author} {\bibfnamefont {D.~L.}\ \bibnamefont {Campbell}},
  \bibinfo {author} {\bibfnamefont {R.~M.}\ \bibnamefont {Price}}, \bibinfo
  {author} {\bibfnamefont {S.}~\bibnamefont {De}},\ and\ \bibinfo {author}
  {\bibfnamefont {I.}~\bibnamefont {Spielman}},\ }\bibfield  {title} {\bibinfo
  {title} {Optimally focused cold atom systems obtained using density-density
  correlations},\ }\href@noop {} {\bibfield  {journal} {\bibinfo  {journal}
  {Review of Scientific Instruments}\ }\textbf {\bibinfo {volume} {85}},\
  \bibinfo {pages} {013110} (\bibinfo {year} {2014})}\BibitemShut {NoStop}%
\bibitem [{\citenamefont {Barenghi}\ and\ \citenamefont
  {Parker}(2016)}]{barenghi2016primer}%
  \BibitemOpen
  \bibfield  {author} {\bibinfo {author} {\bibfnamefont {C.~F.}\ \bibnamefont
  {Barenghi}}\ and\ \bibinfo {author} {\bibfnamefont {N.~G.}\ \bibnamefont
  {Parker}},\ }\href@noop {} {\emph {\bibinfo {title} {A primer on quantum
  fluids}}},\ \bibinfo {number} {arXiv: 1605.09580}\ (\bibinfo  {publisher}
  {Springer},\ \bibinfo {year} {2016})\BibitemShut {NoStop}%
\bibitem [{\citenamefont {Menotti}\ and\ \citenamefont
  {Stringari}(2002)}]{PhysRevA.66.043610}%
  \BibitemOpen
  \bibfield  {author} {\bibinfo {author} {\bibfnamefont {C.}~\bibnamefont
  {Menotti}}\ and\ \bibinfo {author} {\bibfnamefont {S.}~\bibnamefont
  {Stringari}},\ }\bibfield  {title} {\bibinfo {title} {Collective oscillations
  of a one-dimensional trapped bose-einstein gas},\ }\href
  {https://doi.org/10.1103/PhysRevA.66.043610} {\bibfield  {journal} {\bibinfo
  {journal} {Phys. Rev. A}\ }\textbf {\bibinfo {volume} {66}},\ \bibinfo
  {pages} {043610} (\bibinfo {year} {2002})}\BibitemShut {NoStop}%
\bibitem [{\citenamefont {Baym}\ and\ \citenamefont
  {Pethick}(1996)}]{PhysRevLett.76.6}%
  \BibitemOpen
  \bibfield  {author} {\bibinfo {author} {\bibfnamefont {G.}~\bibnamefont
  {Baym}}\ and\ \bibinfo {author} {\bibfnamefont {C.~J.}\ \bibnamefont
  {Pethick}},\ }\bibfield  {title} {\bibinfo {title} {Ground-state properties
  of magnetically trapped bose-condensed rubidium gas},\ }\href@noop {}
  {\bibfield  {journal} {\bibinfo  {journal} {Phys. Rev. Lett.}\ }\textbf
  {\bibinfo {volume} {76}},\ \bibinfo {pages} {6} (\bibinfo {year}
  {1996})}\BibitemShut {NoStop}%
\bibitem [{\citenamefont {Thompson}\ \emph {et~al.}(2013)\citenamefont
  {Thompson}, \citenamefont {Bagnato}, \citenamefont {Telles}, \citenamefont
  {Caracanhas}, \citenamefont {Dos~Santos},\ and\ \citenamefont
  {Bagnato}}]{thompson2013evidence}%
  \BibitemOpen
  \bibfield  {author} {\bibinfo {author} {\bibfnamefont {K.~J.}\ \bibnamefont
  {Thompson}}, \bibinfo {author} {\bibfnamefont {G.~G.}\ \bibnamefont
  {Bagnato}}, \bibinfo {author} {\bibfnamefont {G.~D.}\ \bibnamefont {Telles}},
  \bibinfo {author} {\bibfnamefont {M.~A.}\ \bibnamefont {Caracanhas}},
  \bibinfo {author} {\bibfnamefont {F.~E.~A.}\ \bibnamefont {Dos~Santos}},\
  and\ \bibinfo {author} {\bibfnamefont {V.~S.}\ \bibnamefont {Bagnato}},\
  }\bibfield  {title} {\bibinfo {title} {Evidence of power law behavior in the
  momentum distribution of a turbulent trapped bose--einstein condensate},\
  }\href@noop {} {\bibfield  {journal} {\bibinfo  {journal} {Laser Phys.
  Lett.}\ }\textbf {\bibinfo {volume} {11}},\ \bibinfo {pages} {015501}
  (\bibinfo {year} {2013})}\BibitemShut {NoStop}%
\end{thebibliography}%
\end{document}